%% file: main.tex
\documentclass[sigconf]{acmart}

\copyrightyear{2026}
\acmYear{2026}
\setcopyright{cc}
\setcctype{by}
\acmConference[ICSE '26]{2026 IEEE/ACM 48th International Conference on Software Engineering}{April 12--18, 2026}{Rio de Janeiro, Brazil}
\acmBooktitle{2026 IEEE/ACM 48th International Conference on Software Engineering (ICSE '26), April 12--18, 2026, Rio de Janeiro, Brazil}
\acmPrice{}
\acmDOI{10.1145/3744916.3773113}
\acmISBN{979-8-4007-2025-3/2026/04}

\usepackage{amsfonts}
\usepackage{algorithmic}
\usepackage{textcomp}
\usepackage{xcolor}
\usepackage{framed}
\usepackage{fontawesome}
\usepackage{listings}

\usepackage{amssymb}
\usepackage{multirow}
\usepackage{bbding} 
\usepackage{hyperref}
\usepackage{color}
\usepackage{cleveref}
\usepackage{xspace}
\usepackage{pifont}

\DeclareRobustCommand{\responsebox}[2][gray!20]{%
  \par\medskip\noindent
  \colorbox{#1}{%
    \begin{minipage}{\dimexpr\linewidth-2\fboxsep\relax}
      \vspace{2pt}%
      #2%
      \vspace{2pt}%
    \end{minipage}%
  }%
  \par\medskip
}

\newcommand{\wb}[1]{#1}

\newcommand{\tool}{\textit{IntelliRadar}\xspace}

\begin{document}

\title{IntelliRadar: A Comprehensive Platform to Pinpoint Malicious Packages Information from Cyber Intelligence}

\author{Wenbo Guo}
\email{honywenair@gmail.com}
\affiliation{%
  \institution{Nanyang Technological University}
  \country{Singapore}
}

\author{Chengwei Liu}
\authornote{Corresponding author.}
\email{chengwei.liu@ntu.edu.sg}
\affiliation{%
  \institution{Nanyang Technological University}
  \country{Singapore}
}
\authornote{also with China-Singapore International Joint Research Institute (CSIJRI), Guangzhou, China}

\author{Limin Wang}
\email{wanglimin@smail.nju.edu.cn}
\affiliation{%
  \institution{Nanjing University}
  \country{China}
}

\author{Yiran Zhang}
\email{yiran002@e.ntu.edu.sg}
\affiliation{%
  \institution{Nanyang Technological University}
  \country{Singapore}
}

\author{Jiahui Wu}
\email{jiahui004@e.ntu.edu.sg}
\affiliation{%
  \institution{Nanyang Technological University}
  \country{Singapore}
}

\author{Zhengzi Xu}
\email{z.xu@imperial.ac.uk}
\affiliation{%
  \institution{Imperial Global Singapore}
  \country{Singapore}
}

\author{Yang Liu}
\email{yangliu@ntu.edu.sg}
\affiliation{%
  \institution{Nanyang Technological University}
  \country{Singapore}
}
\authornotemark[2]

\renewcommand{\shortauthors}{Guo, et al.}




\input{body/abstract}

\maketitle

\input{body/intro}

\input{body/background}
\input{body/approach}
\input{body/experiments}
\input{body/discussion}

\input{body/related}
\input{body/conclusion}

\bibliographystyle{ACM-Reference-Format}
\bibliography{ref}

\end{document}

%% file: body/abstract.tex
\begin{abstract}

Malicious packages in public registries pose serious threats to software supply chain security. While current software component analysis (SCA) tools rely on databases like OSV and Snyk to detect these threats, these databases suffer from delayed updates and incomplete coverage. However, they miss intelligence from unstructured sources like social media and developer forums, where new threats are often first reported. This delay extends the lifecycle of malicious packages and increases risks for downstream users.

To address this, we developed a novel and comprehensive approach to construct a platform \tool to 
collect disclosed malicious package names from unstructured web content. Specifically, by exhaustively searching and snowballing the public sources of malicious package names, and incorporating large language models (LLMs) with domain-specialized Least to Most prompts, \tool ensures comprehensive collection of historical and current disclosed malicious package names from diverse unstructured sources.
As a result, we constructed a comprehensive malicious package database containing 34,313 malicious NPM and PyPI package names. Our evaluation shows that \tool achieves high performance (97.91\% precision) on malicious package intelligence extraction. Compared to existing databases, \tool identifies 7,542 more malicious package names than OSV and 12,684 more than Snyk. Furthermore, 76.6\% of NPM components and 70.3\% of PyPI components in \tool were collected earlier than in Snyk's database. \tool is also more cost-efficient, with a cost of \$0.003 per piece of malicious package intelligence and only \$7 per month for continuous monitoring. Furthermore, we identified and received confirmation for 1,981 malicious packages in downstream package manager mirror registries through the \tool.

\end{abstract}

%% file: body/intro.tex
\section{Introduction}

With the widespread global use of open-source software, malicious actors have discovered an effective means to disseminate malicious code through open-source platforms. Especially, TPL registries, such as NPM and PyPI, have become disaster areas of malicious code, in which attackers deliberately upload packages, embedded with malicious code, and induce downstream users to include them as dependencies. 
These malicious packages often carry viruses, Trojans, ransomware, etc.~\cite{npm_remote_trojans,npm_stole_pwd,pypi_malicious_trojans}, and stealthily infiltrate user systems by masquerading as regular libraries or tools. Unlike unintentional vulnerabilities, these are deliberately created by attackers with malicious intent.

To address the threat of malicious packages, academic and industrial researchers have worked to prevent their spread by investigating malicious package types, taxonomies~\cite{guo2023empirical,zhou2024large,zhou2024oss}, and attack surfaces~\cite{okafor2022sok,ladisa2023journey,ladisa2023sok}, as well as developing detection tools~\cite{huang2024donapi,li2023malwukong,vu2023bad,zahan2024shifting}; simultaneously, software composition analysis tools like Snyk~\cite{snykweb}, BlackDuck~\cite{BlackDuckweb}, OWASP Dependency Check~\cite{OWASPweb}, and Dependabot~\cite{dependabotweb} with security databases are used to identify malicious dependencies, relying on platforms such as GitHub Advisory~\cite{advisory_database_ref}, NVD~\cite{nvdweb}, and OSV~\cite{osv_database_ref} for updates. 
However, despite the abundance of detection methods, the malicious packages discovered by these researchers are typically disclosed through scattered unstructured web pages, making it difficult for this information to effectively reach downstream users and developers. Consequently, many identified malicious packages still remain on mirror servers.
Up to 72.4\% of malicious PyPI packages remain on mirror servers after being reported~\cite{guo2023empirical}, with the \textit{colorwed} package~\cite{snyk_colorwed} being downloaded over 1,000 times after identification as malicious, revealing both a lack of security awareness and increased risk as these packages become known to potential attackers during this period.

Therefore, we aim to mitigate these information delays for malicious packages in this paper. Specifically, as detailed by the motivating example in~\Cref{subsec:motivating}, the newly identified malicious packages from researchers, companies, or organizations are usually reported on various public channels, such as personal blogs, tweets, or news on reputed platforms. To address this gap, we aim to establish a comprehensive and public-available intelligence platform to automatically collect, process, and identify malicious packages in time.

\begin{figure*}[htbp]
  \centering
  \vspace{-2mm}
  \includegraphics[scale=0.45]{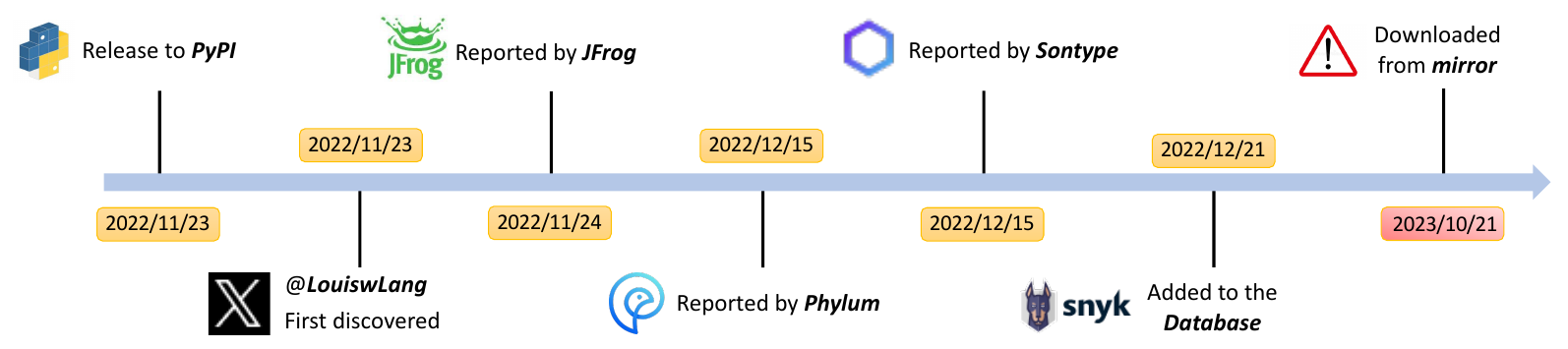}
  \caption{The Timeline of Intelligence Reporting of the Malicious Package \textit{colorwed}}
  \label{fig:motivation}
  \vspace{-1mm}
\end{figure*}

To achieve this, we face the following challenges: 
\textbf{1) Intelligence Sources.} To ensure the intelligence timeliness, we should timely capture first-hand intelligence when they are posted. Therefore, a comprehensive list of intelligence sources should be monitored.
\textbf{2) Key Information Extraction.} Since different reporters formalize intelligence in various ways, 
%such as titles, free-text paragraphs, table of lists, or even code, 
it is still difficult to accurately retrieve the key information of malicious packages, such as package names, corresponding versions, and additional key information (i.e., types and behavior patterns) from intelligence. 
{\textbf{3) Trustworthy of Intelligence.} The intelligence publicly reported online could also be unreliable. 
%It is possible that some intelligence accidentally or deliberately contains inaccurate information, such as fake news and tempering in re-posts. 
Thus, it is also non-trivial to accurately identify inaccurate information before it reaches downstream users.}

To fill these gaps, we propose a comprehensive approach \tool to construct a platform to pinpoint the coverage, timeliness, and accuracy of automated collection and processing of malicious package intelligence. 
Specifically, 
for challenge 1), starting from existing malicious package reports, we conducted a thorough exploration of intelligence sources by summarizing domain-specific keywords and snowballing results from search engines, to identify as many intelligence sources as possible. 
%For challenge 2), we incorporate the large language models (LLMs) to extract the key information of malicious packages precisely. Specifically, to handle the potential inaccuracy (i.e., hallucination~\cite{huang2023surveyhallucinationlargelanguage}), we introduced a multi-faceted approach to prompt LLMs with domain knowledge of malicious packages.
For challenge 2), we incorporate the large language models (LLMs) to extract the key information of malicious packages precisely. Unlike general cyber threat intelligence approaches such as CTIKG~\cite{huang2024ctikg} and SecBERT~\cite{host2023constructing}, our task requires distinguishing malicious package names from benign packages in the same text and performing ecosystem-specific semantic understanding. To handle potential LLM inaccuracy (i.e., hallucination~\cite{huang2023surveyhallucinationlargelanguage}), we introduced a multi-faceted approach to prompt LLMs with domain knowledge of malicious packages.
For challenge 3), we conduct an empirical analysis on conflicted information among intelligence, and based on their correctness, we introduce a voting mechanism based on recency to cross-validate intelligence from different sources, for the same malicious packages. 

Our experiment and analysis show that 
1) Our method collected over 34,313 malicious package names from 24 sources, achieving a clearly higher coverage of malicious packages compared to existing databases (i.e., 12,648 and 7,542 package names are identified as missing in the Snyk Database and OSV database, respectively).
2) Our approach demonstrates exceptional effectiveness in processing and extracting critical information from online threat intelligence sources, achieving 96.4\% recall in keyword-based text filtering. By introducing CoT reasoning and few-shot techniques, our model attains an F1 of 94.87\%, surpassing other LLMs and prompting-based methods.
3) Our intelligence platform can effectively discover the intelligence of malicious packages earlier than existing platforms. More than 76.6\% of NPM and 70.3\% of PyPI malicious package names were discovered earlier than Snyk, with 4,711 malicious packages discovered earlier than OSV.
4) Our approach is highly cost-efficient: text filtering reduces consumption by up to 58.4\%, and with a monthly cost of \$7, each malicious package intelligence extraction costs only \$0.003 via LLM-based analysis.
5) GitHub and Phylum serve as the primary intelligence sources, documenting 45.85\% of NPM and 34.8\% of PyPI intelligence, respectively. GitHub, Phylum, Sonatype, OSV, Checkmarx, and Medium collectively account for more than 70\% of intelligence in both ecosystems.

We summarize the main contributions as follows:

\noindent $\bullet$ We proposed \tool, a comprehensive LLM-based SSC intelligence analysis platform for the complete and in-time collection of malicious package intelligence, achieving an F1-score of 94.87\%.
    
\noindent $\bullet$ We constructed a comprehensive and human-validated dataset containing intelligence on 34,313 malicious package names, establishing the largest database for PyPI and NPM package managers to date, which is publicly accessible through our website~\cite{website}.

\noindent $\bullet$ Our approach demonstrates excellent cost-efficiency, with \tool requiring only \$7 monthly for monitoring all relevant web pages, and each piece of intelligence costing \$0.003 to identify.

\noindent $\bullet$ We reported intelligence on over 1,981 malicious packages to downstream mirror maintainers, significantly contributing to the security of the open-source ecosystem.

Our research adheres to the following ethical principles: (a) We strictly follow website terms of service and robots.txt protocols during data collection; 
(b) We only collect and process publicly available information, with no attempt to access restricted data.

%% file: body/background.tex
\section{Motivation}
\label{subsec:motivating}

Through our analysis of existing malicious package intelligence platforms, we discovered two critical challenges in the current malicious package intelligence ecosystem: (1) delays in threat information propagation and (2) incompleteness of authoritative databases.

\noindent $\bullet$ \textbf{Delay in Intelligence Propagation} In the propagation chain of malicious packages, significant delays exist between initial discovery and eventual inclusion in SSC security databases. Research by Jacobs et al.\cite{10.1093/cybsec/tyaa015} indicates that attackers can exploit such information propagation delays to compromise downstream users' software security. An typical motivating example is illustrated in \Cref{fig:motivation}, the propagation timeline of the malicious PyPI package \textit{colorwed} demonstrates this issue: while the package was discovered and reported on social media\cite{colorwed_twitter} on the same day it was uploaded to the PyPI registry (December 23, 2022), and subsequently reported by multiple security companies (JFrog on November 24\cite{colorwed_jfrog}, Phylum and Sontype on December 15~\cite{colorwed_phylum,colorwed_sonatype}), it took approximately one month (December 21) before being included in software composition analysis tools like Snyk's database. 
%This delay posed serious security risks, as the package was downloaded over 1,000 times during this period. 
%More notably, even after its removal from the official registry, download records from external PyPI mirrors continued until October 21, 2023. 
This case clearly demonstrates the critical timing deficiencies in current malicious package intelligence propagation mechanisms.

\noindent $\bullet$ \textbf{Incompleteness of Authoritative Databases} Currently, malicious package intelligence across mainstream SSC security platforms exhibits significant limitations. While some platforms (e.g., Snyk, OSV, and GitHub Advisory) provide structured databases of malicious packages, their coverage of the entire ecosystem's malicious packages remains limited. Concurrently, a substantial volume of malicious package intelligence is disseminated through unstructured formats such as blog posts, security reports, and technical articles. These unstructured sources often contain rich, detailed information about malicious packages, including their behaviors. However, the unstructured nature of this information impedes automated processing. LLMs offer a promising solution to this challenge by extracting valuable intelligence from these natural language sources through their contextual understanding capabilities.

%% file: body/approach.tex
\section{Approach}

\begin{figure*}[]
  \centering
  \includegraphics[scale=0.4]{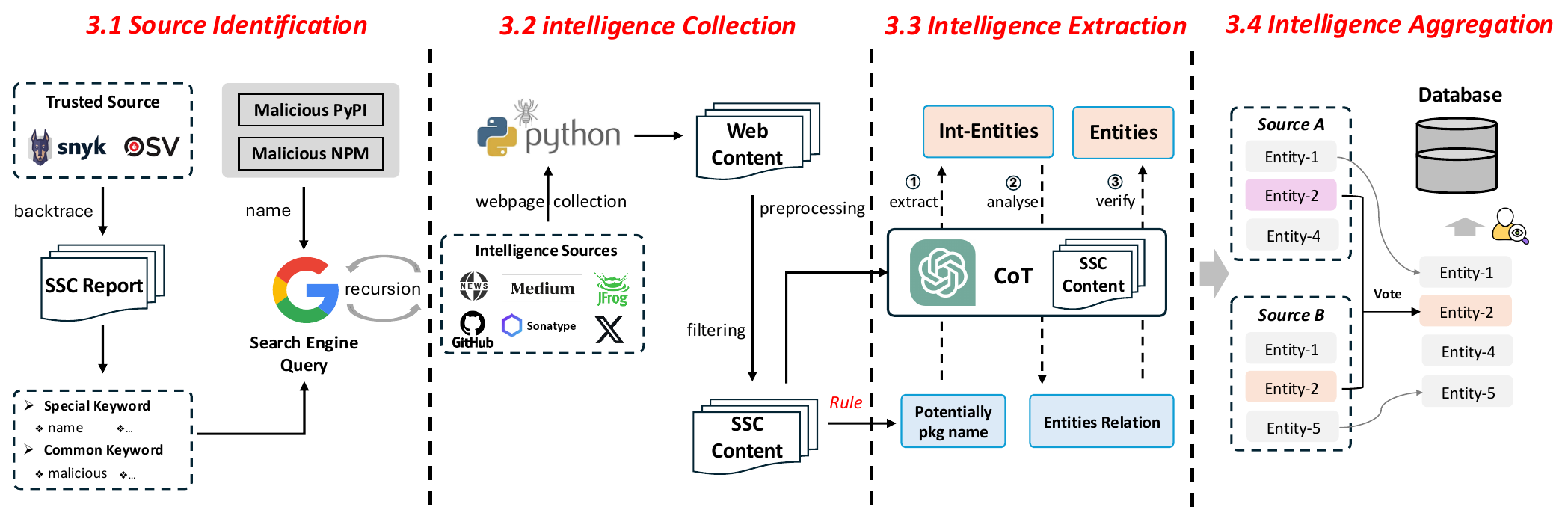}
  \caption{Workflow of the \tool}
  \label{fig:framework}
\end{figure*}

We developed a framework named \tool for collecting, processing, and aggregating open-source malicious package intelligence. As shown in Figure \ref{fig:framework}, our method contains four parts. 
{1) \tool first identify and collect intelligence sources that are highly relevant to the exposure of packages by greedy snowballing searching, after that, 2) it collects all posted intelligences from each source, and extract webpages that possibly contain malicious package related information. Then, 3) \tool employs LLMs with well-designed Chain of Thoughts (CoT) to precisely extract relevant entities and relationships, and based on them, 4) \tool further aggregates these relevant details from different intelligences and derive the complete malicious intelligence database.}

\subsection{Intelligence Source Identification}
\label{subsec:collection}

Identification of the intelligence source is fundamental for intelligence collection. 
{To ensure the completeness of intelligence collection, we first}
% We 
designed a recursive process to identify open source malicious package intelligence sources
{by 3 major steps}
:
\ding{172} keyword selection, \ding{173}  intelligence source identification, and \ding{174} recursive expansion of intelligence sources. 
{Specifically, as presented in~\Cref{fig:framework}, we first construct a comprehensive keyword set to search malicious package related intelligence on the open Internet, and based on the results, we sort out the sources (e.g., websites, accounts, and forums, etc) that constantly post malicious package information, as the targeted package intelligence sources. Moreover, to further ensure that no major sources, especially those post first-hand intelligences, missed, we further introduce a greedy strategy to recursively identify possible intelligence sources documented in identified intelligences, till no new intelligence sources emerges.}

\subsubsection{Keyword Selection.} We first prepare the keywords for searching of existing malicious package related intelligence on the open Internet.
To ensure the coverage and diversity of searched intelligence, we follow a rigorous strategy to extract keywords that are commonly used in existing malicious package reports.

To acquire the existing malicious package reports, we first collect the malicious package list (including 2,351 PyPI and 1,984 NPM packages) from the existing well-known dataset of malicious packages~\cite{ohm2020backstabber}, and cross-map it with Snyk and OSV databases~\cite{osv_database_ref,snyk_database_ref}, which are well verified with reference links, to collect all reference intelligence reports. Based on this corpus, we then select the most relevant keywords that are commonly used in these intelligence reports. Specifically, we collect two groups of keywords, 

\noindent $\bullet$ \textit{Common Keywords}, we aim to identify the keywords that are most commonly used in these malicious intelligence reports. To this end, we apply the Latent Dirichlet Allocation (LDA) model~\cite{tong2016text} to cluster the text topics from the existing reports. After excluding stopwords and irrelevant common words, we selected 17 common keywords that are related to malicious packages out of the top 100 clustered topics.

\noindent $\bullet$ \textit{Special Keywords}, apart from these common keywords, there could also be certain words that are only common in specific report. To this end, we adopt the Term Frequency-Inverse Document Frequency (TF-IDF) algorithm~\cite{qaiser2018text} to identify the top 10 most frequent keywords that are not included in the common keywords in each report. After excluding irrelevant words, we aggregate these words as the special keywords. Interestingly, most of these selected special keywords are the names of malicious packages and their specific attributes, such as execution and control. 
%\cw{can you put 2 or 3 example words here? @wenbo}
The detailed list of keywords are presented in~\Cref{table:keywords}.

\begin{table}[]
\centering
\footnotesize 
\caption{Common and Special Keywords}
\label{table:keywords}
\begin{tabular}{p{3cm}p{5cm}}
\toprule
Type & Keywords \\
\midrule
Special Keywords & <Package name>, <Specific attributes> \\
\addlinespace
Common Keywords
  & package, security, malicious, attacker,
    account, user, registry, code, software,
    github, malware, repository, dataset,
    infected, script, vulnerability, workflow \\
\bottomrule
\end{tabular}
\end{table}

\subsubsection{Intelligence Source Identification.} Based on these keywords, we then conduct a comprehensive searching for malicious package intelligence on the open Internet.

Specifically, we conduct targeted searches through the Google Custom Search API with special keywords configured as mandatory conditions and common keywords configured as optional parameters. Considering that 99.3\% of searches yielding no more than 100 records, for each search, we only take the first 100 records as its result. After deduplication, these pages yielded 7,330 unique webpage links spanning 2,412 distinct web domains. These web domains are considered to be possible sources of malicious intelligence sources.

% \noindent \textbf{Intelligence Source Search:} Based on these keywords, we incorporate a recursive analysis to greedy expand and identify possible sources of malicious package intelligence. We conduct targeted searches through the Google Custom Search API using special keywords as mandatory conditions and common keywords as optional parameters. Ultimately, we collected 22,382 result pages, with 99.3\% of searches yielding fewer than 100 records. After deduplication, these pages yielded 7,330 unique webpage links spanning 2,412 distinct domains.
%We notice that all most all search (99.3\%) yield only less than 100 records in their results. Thus, to avoid too many irrelevant records to be included for some packages that have a common name (i.e., Searching \textit{faq} can yields 121 records in total), we only take the first 100 records from the results of each search. After searching for existing 2,351 PyPI and 1,984 NPM malicious packages~\cite{ohm2020backstabber}, we collected 22,382 result pages. After deduplication, these pages yielded 7,330 unique webpage links spanning 2,400 distinct domains.

% \noindent \textbf{ Intelligence Source Identification:} 
Since sources with only few historical reports of malicious packages are not necessarily to monitor, we further filter out the web domains that are with low frequencies in our search results. 
% To optimize the cost-effectiveness in our intelligence collection, 
% we further filter out 
% we investigate the frequencies of these web domains in our search results. The higher frequency indicates that sources consistently publish information about multiple malicious packages. We verify through sampling that low-frequency domains lack relevance to software supply chain security.
%We first perform a statistical analysis of the frequency distribution of domains, then validate low-frequency domains through sampling, and finally perform a systematic manual evaluation of high-frequency domains.
Specifically, Our statistics revealed that 80.2\% of web domains have only one posts, and only 228 web domains (9.5\%) had ever published over 10 posts in our search results. After we manually inspected the content of these posts, we found that only 24 web domains have published malicious package related information in their historical posts, and they are ultimately identified as the intelligence sources that specialize in SSC security and malicious software analysis in this step for further collection and monitoring.

\subsubsection{Recursive Expansion of Intelligence Sources.} 
Moreover, to ensure that our search did not miss necessary intelligence sources, we then conduct a recursive analysis on the identified intelligence sources, aiming to explore new intelligence sources, till no new source emerges. Specifically, for each of the 24 intelligence sources, we extracted all external links from the web contents we collected in the Google search results. After inspecting the 10,326 newly identified external links, we found that 97\% of these links are actually from the 24 existing intelligence sources that we have already identified, and the remaining 3\% are irrelevant advertising links.

\begin{figure}[]
  \centering
  \includegraphics[width=0.87\linewidth]{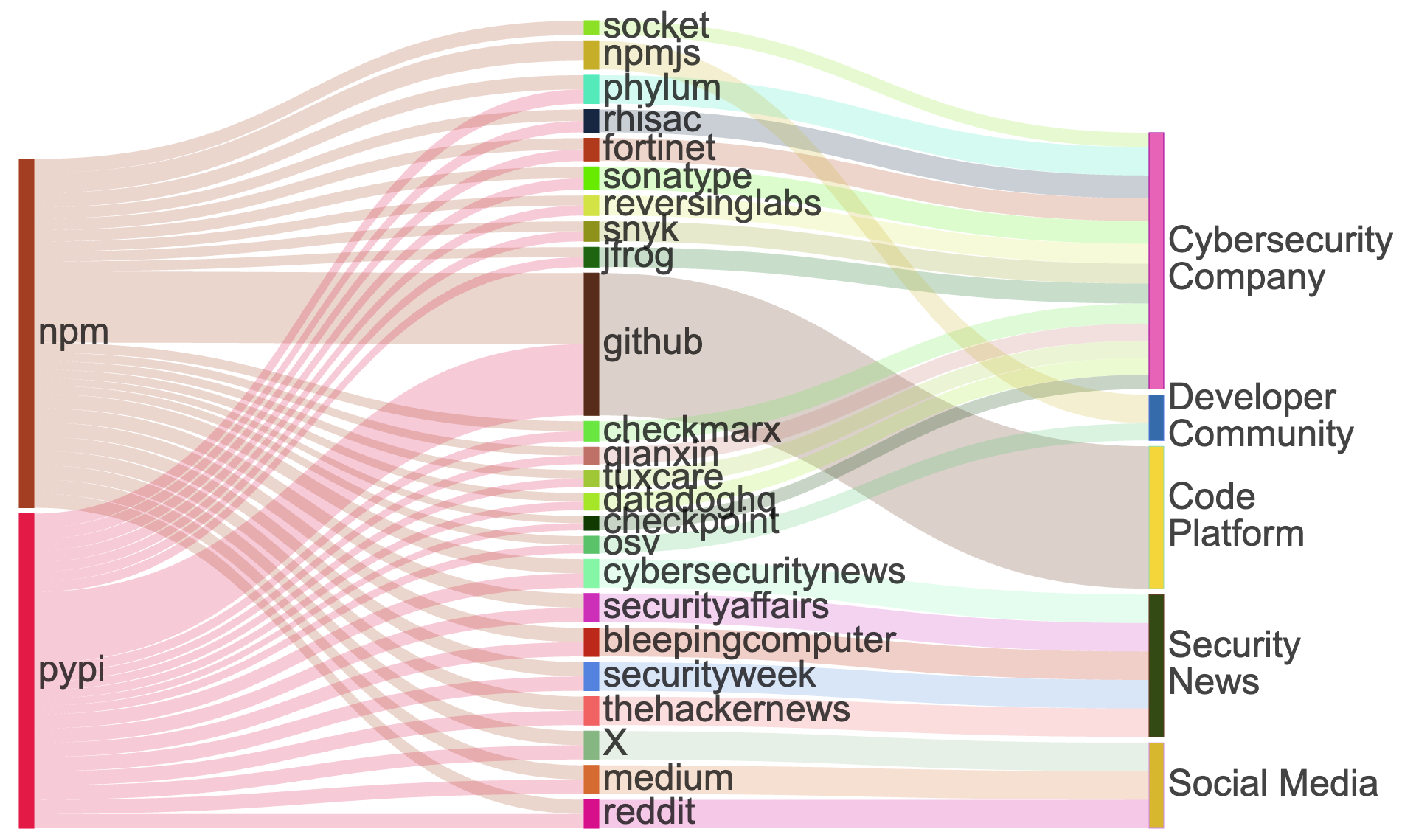}
  \caption{Sources and classification of intelligence sources}
  \label{source_distribution}
\end{figure}

\subsection{Intelligence Collection}

After identifying intelligence sources, we collected all webpage content for historical intelligence extraction and continuous monitoring. As shown in Figure~\ref{source_distribution}, we classified these sources into five functional categories. Each category employs distinct structures for SSC intelligence distribution: Cybersecurity companies utilize specialized second- or third-level domains, Security News sites implement keyword indexing systems, while developer communities, code platforms, and social networks distribute intelligence through specific accounts. These structural patterns enabled targeted collection from relevant subdomains and accounts, eliminating the need for comprehensive site-wide crawling. However, even within these targeted SSC-related intelligence sources, many webpages still contain content irrelevant to malicious package intelligence. 
%To overcome this challenge, we leveraged the common keywords extracted in ~\Cref{subsec:collection} as filtering criteria, thereby significantly reducing the processing burden of subsequent intelligence extraction steps while maintaining comprehensive coverage of relevant information. 
To overcome this challenge, we established clear inclusion and exclusion criteria for filtering relevant webpages. Our filtering employs a two-tier approach: ecosystem-specific keywords (e.g., "pypi," "npm") as mandatory criteria, and common security keywords from ~\Cref{table:keywords} (e.g., "malicious," "security," "package") as optional filters. This mechanism significantly reduces computational overhead while maintaining comprehensive coverage. To collect webpage content, we developed specialized HTML parsers targeting information-rich elements (tables, lists, paragraphs, code snippets, headings) and iframes containing malicious package details and IOCs.

\subsection{Malicious Package Intelligence Extraction}
\label{subsec:extraction}

\begin{figure*}[]
  \centering
  \includegraphics[scale=0.72]{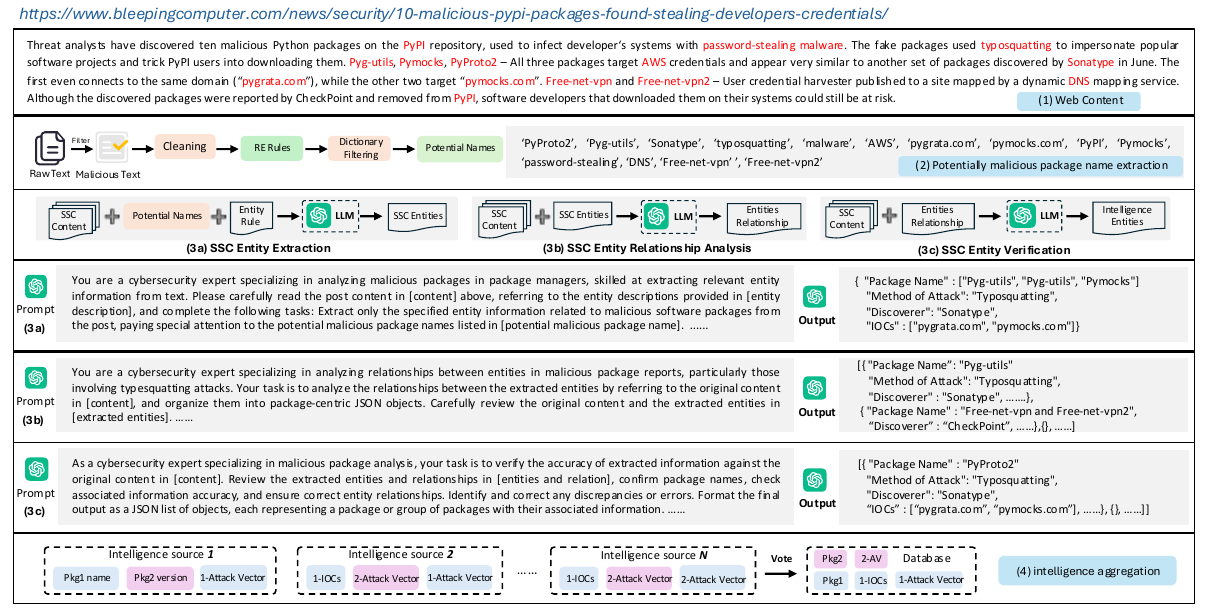}
  \caption{\wb{Entity and Relationship Analysis Using the CoT Prompts}}
  \label{fig:gpt_analysis}
\end{figure*}

{As shown in Figure \ref{fig:gpt_analysis}~(1), the web content often includes a significant amount of dispersed information related to malicious packages, which is highlighted in yellow for clarity and emphasis. To efficiently extract the key intelligence from these texts, we employed LLM to enhance the information extraction process~\cite{min2023recent, wang2023gpt}. 
However, LLM still faces issues like hallucination, limiting extraction accuracy~\cite{ling2023improving}. To address this, we first preprocess the web content to extract potential names of malicious packages. Then, we feed both the extracted potential malicious package information and the web content together into LLM to extract more detailed and accurate malicious package information~\cite{martino2023knowledge}. }

{The potential malicious package name extraction process is illustrated in Figure \ref{fig:gpt_analysis}~(2). We have observed that malicious packages on PyPI and NPM often try to deceive downstream users into downloading them by creating package names that are similar to benign packages, such as through \textit{typosquatting}~\cite{tahir2018s, vu2020typosquatting} creating package names with minor letter variations. As a result, the names of these malicious packages often aren't real words found in English dictionaries.
Based on this observation, we designed a two-step extraction pipeline. First, we employ the regular expression r"(?:@[a-zA-Z0-9-]+/)?[a-zA-Z0-9][a-zA-Z0-9.\_-]+" to extract strings matching package naming conventions from webpage content. 
%This pattern captures NPM scoped packages (prefixed with @), NPM regular packages, and PyPI packages according to their standard naming specifications. 
Subsequently, we apply filtering operations to the extracted candidate package names: removing duplicates to eliminate redundancy, stripping trailing punctuation marks such as periods and commas, and filtering out common English vocabulary and stopwords, thereby retaining potential malicious package names.

To analyze malicious package intelligence, we employ CoT reasoning with three components: entity extraction, relationship analysis, and verification. This decomposition aligns with how security analysts process intelligence: extracting discrete data, revealing patterns and connections, then ensuring integrity.

%This structured approach allows us to systematically process complex security intelligence while maintaining analytical rigor throughout each phase of analysis.

\noindent\textbf{(1) Entity Extraction (Fig~\ref{fig:gpt_analysis}a)}:
Identify and extract key entities related to malicious packages. The prompt design consists of six components: 1) task description, 2) intelligence content, 3) 9 entity types including Package Name, Version, Date of Discovery, Repository URL, Method of Attack, Discoverer, Impacted Systems, Attack Vector, and IOCs, 4) potential package names from previous steps, 5) common malicious naming patterns (typosquatting and misspelling), and 6) few-shot examples. The LLM employs a Chain-of-Thought approach with progressive difficulty: first confirming malicious package names, then extracting observable information (versions, dates, URLs), followed by inferring complex semantics (attack methods, IOCs), and finally grouping packages with shared characteristics, producing structured JSON output containing the nine entity types.

% Identify and extract key entities related to malicious packages. The prompt includes: 1) task description, 2) intelligence content, 3) 9 entity types (Package Name, Version, Discovery Date, Repository URL, Attack Method, Discoverer, Impacted Systems, Attack Vector, IOCs), 4) potential package names, 5) malicious naming patterns (typosquatting, misspelling), and 6) few-shot examples. The LLM uses Chain-of-Thought with progressive difficulty: confirming package names, extracting observable information (versions, dates, URLs), inferring complex semantics (attack methods, IOCs), then grouping packages, ultimately producing structured JSON output.

\noindent\textbf{(2) Entity Relationship Analysis (Fig~\ref{fig:gpt_analysis}b)}:
% Analyze the relationships between extracted entities and organize information centered around packages. We input the entities extracted in the first stage along with the original text, LLM then analyzes the associations between entities, generating a structured JSON output.
Analyze the relationships between extracted entities and organize information centered around packages. The prompt design integrates four components: 1) task description of analyzing semantic relationships between extracted entities, 2) the intelligence content, 3) identified entities from the previous extraction step, and 4) few-shot examples as reference. The LLM analyzes the associations between entities, generating a structured JSON output.

\noindent\textbf{(3) Entity Verification (Fig~\ref{fig:gpt_analysis}c)}:
% This stage verifies the accuracy of the extracted information, ensuring consistency and completeness of the output. By inputting both the original text and the results of entity relationship analysis into the LLM, we enable it to cross-check the accuracy of entities and their relationships against the original text. The LLM then makes corrections as necessary to ensure the final output is reliable and accurate.
This stage verifies the accuracy of the extracted information through LLM-based cross-validation and self-correction. The process inputs both the original intelligence content and the structured entities from the entity relationship analysis step. The LLM performs cross-validation by comparing the extracted entities and their relationships against the original text, identifying inconsistencies or errors, and making necessary corrections to ensure output maintains accuracy and completeness.

During the prompt design phase, we performed iterative refinement based on small-scale performance observations to optimize the LLM extraction process. Based on the experimental results and analysis, we implemented two major prompt enhancements: 1) specifying a particular output structure and providing case studies to guide LLM in extracting the required information more accurately, and 2) updating the prompts with specific examples for correction. These design refinements enabled the LLM to perform the extraction task with reliable precision (as detailed in~\Cref{sec:rq1}).

% \begin{table}[!tb]
% \centering
% \scriptsize
% \setlength{\tabcolsep}{3pt}
% \renewcommand{\arraystretch}{1.2}
% \caption{Entity Extraction: "ef323refefeffe" Package}
% \wb{
% \begin{tabular}{l|p{5.5cm}}
% \hline
% \textbf{Entity Type} & \textbf{Extracted Value} \\
% \hline
% \textbf{Package Name} & ef323refefeffe \\
% \textbf{Package Manager} & PyPI \\
% \textbf{Version} & 1.0 \\
% \textbf{Discovery Date} & 2023-12-26 \\
% \textbf{Repository URL} & \begin{minipage}[t]{5cm}
% \scriptsize\texttt{https://socket.dev/pypi/package/ef323refefeffe\\/files/1.0/tar-gz/ef323refefeffe-1.0/setup.py}
% \end{minipage} \\
% \hline
% \textbf{Method of Attack} & stealing sensitive information, \\ 
% & abusing Discord webhooks \\
% \textbf{Discoverers} & Socket research team \\
% \textbf{Impacted Systems} & Windows \\
% \textbf{Attack Vector} & Exploiting Discord \\
% \hline
% \textbf{IOC Indicators} & \begin{minipage}[t]{5cm}
% \scriptsize\texttt{https://canvas.discord.com/api/webhooks/118942742960\\1710100/JmLvp-Xz...a42Le\_zNGMEa8p5V\_VZxHh4VsEUpzVp6X}
% \end{minipage} \\
% \hline
% Collection Timestamp & 2024-01-09 \\
% \hline
% \multicolumn{2}{l}{\scriptsize Note: Extracted from} \\
% \multicolumn{2}{l}{\scriptsize \href{https://socket.dev/blog/blank-grabber-python-package-steals-info-from-discord-and-telegram}{https://socket.dev/blog/blank-grabber-python-package-steals-info-from-discord-and-telegram}} \\
% \end{tabular}}
% \label{tab:entity_extraction_example}
% \end{table}

\subsection{Intelligence Aggregation}

Through the above methods, we can accurately extract information related to malicious packages from webpages. When multiple sources provide entity information on the same malicious package, we need to aggregate entity information from different sources to ensure comprehensive intelligence gathering. For each malicious package P, we collect a set of entity information $E = \{N, V, F, R, M, D, I, A, C, T\}$, which includes name N, version number V, discovery date F, repository URL R, attack method M, discoverer D, affected system I, attack vector A, IOC indicators C, and and collection timestamp T which represents the webpage publication time, extracted from webpage content during crawling.

Our analysis revealed that while most entity information is consistent across sources, conflicts do exist, particularly in attack methods and version numbers (5.13\%), affected systems (2.45\%), and repository URLs (0.41\%). For these conflicting cases, we observed two patterns: (1) later webpages tend to be more accurate as they often incorporate and reference earlier findings, and (2) information confirmed by multiple sources demonstrates higher reliability. 
 
 %Based on these observations, we employ two strategies for intelligence aggregation to ensure comprehensiveness and accuracy of the information. First, for intelligence from different sources with non-overlapping entity information, we directly merge them. Second, for intelligence from different sources with partially overlapping information, we use a voting mechanism for aggregation. We then add manual verification to ensure the accuracy of the aggregated information. 
 Based on these observations, we employ two aggregation strategies: (1) direct merging for non-overlapping entity information from different sources, and (2) voting mechanism for partially overlapping information. Manual verification ensures aggregated information accuracy.

\textbf{Voting Mechanism}: For each package name $N$ of the malicious package information, vote on fields $V$, $R$, $M$, $D$, $I$, $A$. After excluding $NaN$ values, count the occurrences of each value. Let $E_i$ be the value of entity $E$ in the \textit{i}th intelligence source. Define the voting count function $count(E_i)$ to represent the number of occurrences of value $E_j$. If there is a $\exists E_i, E_j$ such that $count(E_i) = count(E_j)$, select the entry with the largest timestamp $T$: $E_{\text{final}} = \arg\max_{E_i} T_i$. For the discovery date $F$, select the earliest date: $F_{\text{final}} = \min_{i} F_i$. For IOCs $C$, merge all non-duplicate values: $C_{\text{final}} = \bigcup_{i} C_i$. 

Through this voting-based aggregation mechanism, we can effectively aggregate entity information of malicious packages while ensuring the completeness of intelligence. The experimental results in ~\Cref{sec:rq1} validate the effectiveness of our voting-based aggregation approach, demonstrating its ability to accurately resolve conflicts and produce high-quality aggregated intelligence.

%To ensure the accuracy of our final database, we implemented a two-step verification process. First, we verified package existence by validating against official npm and PyPI repository metadata. Second, we confirmed maliciousness through multiple approaches: for npm packages, we used the presence of a "0.0.1-security" version as an indicator of previously flagged malicious packages; for packages that couldn't be verified through this method, we manually reviewed evidence from multiple independent intelligence sources. This cross-source validation approach was particularly effective, as packages confirmed by multiple sources carried stronger evidence of maliciousness, enabling us to build a reliable database of verified malicious packages.

To ensure the accuracy of our final database, we implemented a two-step verification process. First, we verified package existence by validating against official npm and PyPI repository metadata APIs~\cite{npmregistry,pypistats,pypibigquery}, which retain historical package information even after malicious packages are removed, ensuring extracted names correspond to actual packages rather than LLM hallucinations. Second, we confirmed maliciousness through multiple approaches: cross-referencing with established security databases (OSV, Snyk, GitHub Advisory), multi-source corroboration for packages reported by multiple independent sources, and manual verification for remaining packages, which proved efficient due to malicious packages typically being released in batches. Through this multi-layered verification approach, we established a reliable and accurate database of verified malicious packages.

% Through the voting algorithm, when an intelligence entity is confirmed by multiple intelligence sources, we consider its quality and reliability to be higher the more widespread its sources are (see Sections \ref{sec:rq3} and \ref{sec:rq4} for RQ3 and RQ4). This approach enhances the completeness and diversity of intelligence while ensuring the quality and relevance of aggregated information by considering the number of confirmations from different intelligence sources.

%% file: body/experiments.tex
\section{Experiments}

% Malicious packages in SSC follow a lifecycle as described by Guo et al. \cite{guo2023empirical}, consisting of release, distribution, discovery, and removal phases. Our work focuses specifically on the critical window between initial discovery and removal. This window can be further subdivided into the time between first public discovery~(in the wild), discovered by \tool, recorded in OSV/Snyk databases, and eventual removal from registries and mirrors. Reducing delays within this intelligence propagation chain is essential for minimizing downstream risks.

\subsection{Research Questions}
To evaluate the effectiveness of \tool, we designed our experiments around five distinct research questions:

\noindent $\bullet$ \textbf{RQ1~(\textbf{Effectiveness})}: What is the effectiveness of each key component in \tool for intelligence extraction?

\noindent $\bullet$ \textbf{RQ2~(\textbf{Completeness})}: How does \tool compare to existing databases on the completeness of collected SSC Intelligence?

\noindent $\bullet$ \textbf{RQ3~(\textbf{Timeliness})}: How does \tool compare to existing databases in terms of the timeliness of intelligence collection?

\noindent $\bullet$ \textbf{RQ4~(\textbf{Source Distribution})}: How do different intelligence sources contribute to \tool's intelligence collection?

\noindent $\bullet$ \textbf{RQ5~(\textbf{Usability})}: How does \tool's availability contribute to mitigating threat propagation in the downstream software supply chain ecosystem?

Our experimental setup outlines the dataset collected throughout this work. RQ1 examines the effectiveness of key components in our methodology, while RQ2 and RQ3 represent our paper's primary contributions: the completeness and timeliness of malicious package intelligence. RQ4 investigates how different intelligence sources contribute to these core attributes. RQ5 assesses the real-world impact of our approach in protecting downstream ecosystems.

\subsection{Experimental Setup}
Through comprehensive data collection from 24 intelligence sources that were constructed and identified in Section~\ref{subsec:collection}, including both unstructured webpages and structured databases (OSV, Snyk, GitHub Advisory). we collected 50,586 raw webpage texts. 
After filtering, we retained 28,593 webpages containing security intelligence. Through LLM analysis, we found that 11,173 (39.1\%) of these security-related webpages contained no malicious package names. From the remaining webpages, we extracted 35,229 potentially malicious package records.
We first validated these records against official package registries, confirming that 34,982 (99.3\%) packages existed in npm and PyPI registries, while 247 (0.7\%) package names did not exist in these registries, indicating they were not real packages. For accuracy validation, we cross-referenced our dataset with established security databases, which verified 29,485 packages (83.7\%) as malicious packages. The remaining 5,744 packages underwent our multi-source verification process, where packages reported by at least two independent sources were automatically classified as confirmed malicious (3,868 packages, representing 67.5\% of the verification subset). For the remaining 1,876 single-source packages, we conducted manual verification, leveraging the observation that malicious packages are typically distributed in batches and can therefore be verified collectively when referenced in the same intelligence source. This verification identified 960 additional malicious packages, while 916 (2.6\% of the original collection) were classified as benign. After this verification process, our final database contains 34,313 confirmed malicious package names (12,522 PyPI and 21,791 npm packages), achieving a precision rate of 97.4\%.

\subsection{RQ1: Effectiveness}
\label{sec:rq1}

Regarding the effective collection and identification of malicious package intelligence, in this section, we first investigate the contribution of each steps to effectiveness.

\noindent \textbf{LLM Parameter Configuration.} Throughout all experiments in this study, we maintain consistent LLM parameter settings to ensure fair comparison and reproducibility. Specifically, we configure the temperature parameter to 0 to minimize randomness in model outputs, and set top\_p to 0.3 to control the diversity of generated responses. We utilize GPT-4o version gpt-4o-2024-11-20 and GPT-4o-mini version gpt-4o-mini-2024-07-18, both with knowledge cutoff of October 2023. All models are accessed through Azure-provided OpenAI API with API version 2024-02-15-preview. All models and experiments are conducted under this unified parameter configuration.

% In the previous section, we discussed the malicious package data collected throughout our pipeline. In this section, we conducted a series of experiments to evaluate the effectiveness of \tool's key components . 

\noindent \textbf{Validation of keyword text filtering.} To evaluate the effectiveness of keyword text filtering, we randomly selected 250 text samples related to malicious package intelligence and 250 unrelated text samples. The experimental results shown in Table~\ref{tb:keyword_filtering_performance} indicate that the system successfully filtered 88.9\% of irrelevant texts (220/250), with only 30 irrelevant samples misclassified as relevant. Meanwhile, among the 250 relevant samples, only 9 were incorrectly filtered out, resulting in a recall rate of 96.4\%. To further validate our filtering effectiveness at scale, we conducted comprehensive analysis on all 50,586 original webpages using broader security-related terms. This expanded filtering identified 5,893 additional pages, from which manual examination revealed only 27 additional malicious packages (0.08\% of our total collection), confirming that our original approach captured the vast majority of relevant intelligence.

%These errors mainly stemmed from overly broad keyword filter settings leading to false positives, while false negatives were due to inherent limitations of the keyword filtering mechanism. Although some keywords might appear to be related to package manager information on the surface, the actual text content may not involve specific package manager intelligence, but rather cover general information such as news or tool introductions.

%The experimental results showed that the accuracy of keyword filtering reached 92.2\%. Specifically, 220 unrelated samples were successfully filtered out, with only 30 texts incorrectly identified as related to package manager intelligence, successfully filtering 88.9\% of irrelevant texts and improving the efficiency of subsequent analysis tasks. Meanwhile, only 9 relevant intelligence samples were erroneously filtered out. 

\begin{table}[!tb]
\footnotesize
\centering
\setlength{\tabcolsep}{6pt}  % 调整列间距以使表格更紧凑
\renewcommand{\arraystretch}{1.1}
\caption{Accuracy of Keyword Filtering for Malicious Package-Related and Unrelated Texts}
\scalebox{1.0}{
\begin{tabular}{lccccc}
\toprule
\textbf{Metric} & \textbf{Precision} & \textbf{Recall} & \textbf{F1} & \textbf{FN} & \textbf{FP} \\
\midrule
\textbf{Keyword Filtering} & 88.9\% & 96.4\% & 92.5\% & 9 & 30 \\
\bottomrule
\end{tabular}}
\label{tb:keyword_filtering_performance}
\end{table}

% \begin{table}[!tb]
% \centering
% \small
% \setlength{\tabcolsep}{3pt}  % 调整列间距以使表格更紧凑
% \renewcommand{\arraystretch}{1.1}
% \caption{Comparison of \tool and Baseline Models on Entity Extraction and Entity Relationship Alignment}
% \scalebox{1.0}{
% \wb{
% \begin{tabular}{c|ccc|ccc}
% \hline
% \multirow{2}{*}{\textbf{Solution}} & \multicolumn{3}{c|}{\textbf{Entity Extraction}} & \multicolumn{3}{c}{\textbf{Relationship}} \\
% \cline{2-7}
%  & \textbf{Prec.} & \textbf{Recall} & \textbf{F1} & \textbf{Prec.} & \textbf{Recall} & \textbf{F1} \\
% \hline
% \textbf{\tool} & 92.9\% & 86.4\% & 89.5\% & 98.6\% & 86.2\% & 92.0\% \\
% No-potential pkgs & 91.8\% & 81.4\% & 86.3\% & 96.9\% & 80.9\% & 88.2\% \\
% \hline
% \end{tabular}
% }
% }
% \label{tb:comparison}
% \end{table}

\begin{table}[!tb]
\centering
\footnotesize
\setlength{\tabcolsep}{3pt}
\renewcommand{\arraystretch}{1.1}
\caption{Evaluation of Different LLMs and Methods}
\scalebox{1.0}{
\begin{tabular}{c|ccccc}
\hline
\textbf{Model} & \textbf{Precision} & \textbf{Recall} & \textbf{F1} & \textbf{FN} & \textbf{FP} \\
\hline
LLaMA3.3-70B & 95.63\% & 67.46\% & 79.11\% & 232 & 22 \\
LLaMA3.1-70B & 97.25\% & 54.56\% & 69.90\% & 324 & 11 \\
Qwen2.5-72B & 95.20\% & 75.04\% & 83.92\% & 178 & 27 \\
GPT-4o-mini & 91.53\% & 63.67\% & 75.10\% & 259 & 42 \\
\hline
\wb{CTIKG~\cite{huang2024ctikg}} & \wb{53.85\%} & \wb{12.44\%} & \wb{20.22\%} & \wb{591} & \wb{72} \\
\wb{SecBERT~\cite{host2023constructing}} & \wb{0.10\%} & \wb{2.07\%} & \wb{0.20\%} & \wb{661} & \wb{13,340} \\
\hline
\tool$_{few-shot}$ & 95.74\% & 59.89\% & 73.68\% & 286 & 19 \\
\tool$_{CoT}$ & 94.51\% & 91.73\% & 93.10\% & 59 & 38 \\
\textbf{\tool} & \textbf{97.91\%} & \textbf{92.01\%} & \textbf{94.87\%} & \textbf{57} & \textbf{14} \\
\hline
\multicolumn{6}{l}{\small Note: Results based on the final verification stage of LLM analysis.} \
\end{tabular}
}
\label{tab:model_comparison}
\end{table}

\begin{table}[]
\centering
\footnotesize
\caption{Model Performance on Recent Security Reports}
\label{tab:performance_summary}
\renewcommand{\arraystretch}{1.1}
\wb{
\begin{tabular}{@{}lccccc@{}}
\toprule
\textbf{Model} & \textbf{Precision} & \textbf{Recall} & \textbf{F1} & \textbf{FP} & \textbf{FN} \\
\midrule
\textbf{\tool}~(GPT-4o)        & \textbf{99.17\%} & \textbf{88.81\%} & \textbf{93.70\%} & 1  & 15 \\
GPT-4o-mini   & 87.80\% & 80.60\% & 84.05\% & 15 & 26 \\
Qwen2.5-72B   & 97.96\% & 71.64\% & 82.76\% & 2  & 38 \\
LLaMA3.1-70B  & 95.08\% & 86.57\% & 90.62\% & 6  & 18 \\
LLaMA3.3-70B  & 92.19\% & 88.06\% & 90.08\% & 10 & 16 \\
\bottomrule
\end{tabular}}
\end{table}

\noindent \textbf{Validation of entity extraction.}
To evaluate to what extent selecting different prompting strategies, as well as different LLMs, could influence the effectiveness of \tool,
we implement two sets of variants of our approach for comparison: 1) we implement our approach with different prompting strategies, \tool$_{few-shot}$ using only a few shots, \tool$_{CoT}$ implementing only CoT reasoning, to evaluate which contributes more to our complete approach (\textbf{\tool}) that integrates both strategies. 2) We also implement our approach with different LLMs, for instance, LLaMA3.3-70B, LLaMA3.1-70B, Qwen2.5-72B, and GPT-4o-mini, to evaluate the influence of select GPT-4o to our approach. We also compared with existing approaches: CTIKG achieved 20.22\% F1-score due to lack of explicit entity definitions for malicious packages, while SecBERT performed poorly (0.20\% F1) as it incorrectly identified numerous irrelevant text spans, resulting in extremely high false positives.
The evaluation is performed on the 713 ground truth packages in the validation of entity extraction, treating all versions of the same package as a single entity where extraction is considered correct only when both package name and all associated versions match exactly. Our evaluation framework defines three outcomes: correctly extracted packages (TP), missed packages in webpages (FN), and incorrectly extracted package names (FP).
As shown in Table \ref{tab:model_comparison}, the experimental results demonstrate significant variations in the performance of the model in different LLMs and prompting strategies. 
%\textbf{\tool} approach achieves optimal performance with a precision of 97.91\%, and recall of 92.01\%. The low false positive count (14) and false negative count (57) indicate the model's robust ability to accurately identify and extract malicious package information. 
\textbf{\tool} approach achieves optimal performance with a precision of 97.91\%, and recall of 92.01\%. The low FP count (14) indicates minimal incorrect extractions of non-existent packages, benign packages, and malformed data, while the low FN count (57) shows few malicious package entities in the text were missed during extraction, demonstrating the model's robust ability to accurately identify malicious package information.

To address data leakage concerns, we conducted an additional evaluation on 50 security intelligence webpages published after January 1, 2025, ensuring all content postdates the training cutoff of evaluated models. As shown in Table~\ref{tab:performance_summary}, \tool achieves 93.70\% F1-score with only 1 false positive, indicating strong precision on unseen data. These findings align with our evaluation results (Table~\ref{tab:model_comparison}), demonstrating that \tool consistently outperforms baseline models on both seen and unseen data. Beyond our core task of extracting package names and versions, we also evaluated other security-related entities. As shown in Table~\ref{tab:information_extraction_performance}, our method demonstrates good performance on supplementary entities such as IOC (87.10\% F1) and Discoverer (87.32\% F1) extraction.

% To address potential data leakage concerns where the malicious intelligence might have been encountered during the training phase of these models, we conducted an additional evaluation on fresh data. We randomly selected 50 security intelligence webpages published after January 1, 2025, ensuring that all content postdates the training cutoff of the evaluated models. As shown in Table \ref{tab:performance_summary}, \textbf{\tool} achieves the best overall performance with 93.70\% F1-score, with only 1 false positive indicating strong precision maintenance on unseen data. These findings on unseen data align with our evaluation results (Table \ref{tab:model_comparison}), demonstrating that \textbf{\tool} consistently outperforms baseline models regardless of whether the malicious package intelligence was encountered during training. In addition to our core task of extracting malicious package names and versions, we also evaluated our approach on extracting other supplementary security-related entities. As shown in Table~\ref{tab:information_extraction_performance}, while our primary contribution targets package identification, our method also demonstrates good performance on additional entities such as IOC (87.10\% F1) and Discoverer (87.32\% F1) extraction.}

From the perspective of different prompting strategies, our results reveal that the Chain-of-Thought (CoT) approach substantially improves the recall rate from 59.89\% (\tool$_{few-shot}$) to 91.73\% (\tool$_{CoT}$). This improvement can be attributed to CoT's systematic decomposition of the extraction task, enabling a more comprehensive identification of package entities. In particular, while \tool$_{CoT}$ achieves a high recall, its false positive count (38) is higher than \textbf{\tool} (14), demonstrating how the combination of both strategies helps maintain precision while maintaining high recall.

\begin{table}[]
\centering
\footnotesize
\caption{Other Intelligence Extraction Performance on Security Reports}
\label{tab:information_extraction_performance}
\renewcommand{\arraystretch}{1.1}
\wb{
\begin{tabular}{@{}lccc@{}}
\toprule
\textbf{Entity} & \textbf{Precision} & \textbf{Recall} & \textbf{F1} \\
\midrule
Date of Discovery  & \textbf{95.24\%} & 83.33\% & \textbf{88.89\%} \\
Discoverer         & 86.11\% & \textbf{88.57\%} & 87.32\% \\
IOC               & 80.60\% & \textbf{94.74\%} & 87.10\% \\
Method of Attack  & 75.00\% & 78.95\% & 76.92\% \\
Attack Vector     & 72.00\% & 75.00\% & 73.47\% \\
Impacted Systems  & 64.71\% & 68.75\% & 66.67\% \\
Repo URLs         & 53.85\% & 77.78\% & 63.64\% \\
\bottomrule
\end{tabular}}
\end{table}

% \begin{table}[h]
% \small
% \renewcommand{\arraystretch}{1.1}
% \caption{Experimental Results of Entity Aggregation}
% \begin{tabular}{lcccc}
% \hline
% \textbf{} & \textbf{Total Entities} & \textbf{Correct} & \textbf{Incorrect} & \textbf{Acc.} \\
% \hline
% \textbf{\tool} & 200 & 181 & 19 & 90.5\% \\
% \hline
% \multicolumn{5}{l}{\small Note: The 200 entities were extracted from 635 source reports.} \\
% \end{tabular}
% \label{tab:merging-accuracy}
% \end{table}

% To provide context about the aggregation requirements in our dataset, we analyzed the distribution of entities requiring voting-based aggregation. Among all entities, only a small portion needed voting-based aggregation, with Method of Attack and Version showing the highest proportion (5.13\%), followed by Impacted Systems (2.45\%) and Repository URL (0.41\%). 
\noindent \textbf{Validation of intelligence aggregation}
To evaluate the accuracy of our entity aggregation approach, we randomly selected 200 entities that required aggregation, which originated from 635 different intelligence source reports. Each selected entity contained conflicting information that needed to be consolidated. To prove the effectiveness of our voting mechanism, we manually label the ground truth for each of them.
% As shown in Table~\ref{tab:merging-accuracy}, 
Our voting approach achieved an accuracy of 90.5\%, successfully aggregating 181 entities correctly while incorrectly aggregating 19 entities, which confirms the majority are right in the field of malicious package intelligence, and also proves that our simple voting mechanism can fit most of conflicting cases in malicious package intelligence. This reduces the cost of manual inspection.
% indicating the effectiveness of our voting-based intelligence aggregation method.

% From the perspective of different LLMs, the performance of different LLMs vary. Qwen2.5-72B achieves the highest baseline performance with an F1 score of 83.92\%, while LLaMA3.3-70B and LLaMA3.1-70B show a precision recall trade-off pattern. The high precision (97.25\%) but low recall (54.56\%) of LLaMA3.1-70B indicates a conservative extraction pattern, suggesting potential limitations in identifying complex or ambiguous package references. The superior performance of \textbf{\tool} mainly stems from two technical innovations: (1) the integration of structured CoT reasoning with contextual few-shot examples, which enhances the model's comprehension of domain-specific intelligence extraction requirements, and (2) the optimized prompting strategy that effectively balances precision and recall in entity identification. These results empirically validate the effectiveness of combining both CoT reasoning and few-shot learning in specialized intelligence extraction tasks.

\responsebox{
Response to RQ1: The text filtering effectively removes irrelevant content, with \tool outperforming all baseline models in entity extraction and relationship alignment. By combining CoT prompting with few-shot, \tool achieves an F1 score of 94.87\%, surpassing other LLMs.
}

\subsection{RQ2: Completeness}
\label{sec:rq2}

\begin{figure}[]
  \centering
  \includegraphics[scale=0.14]{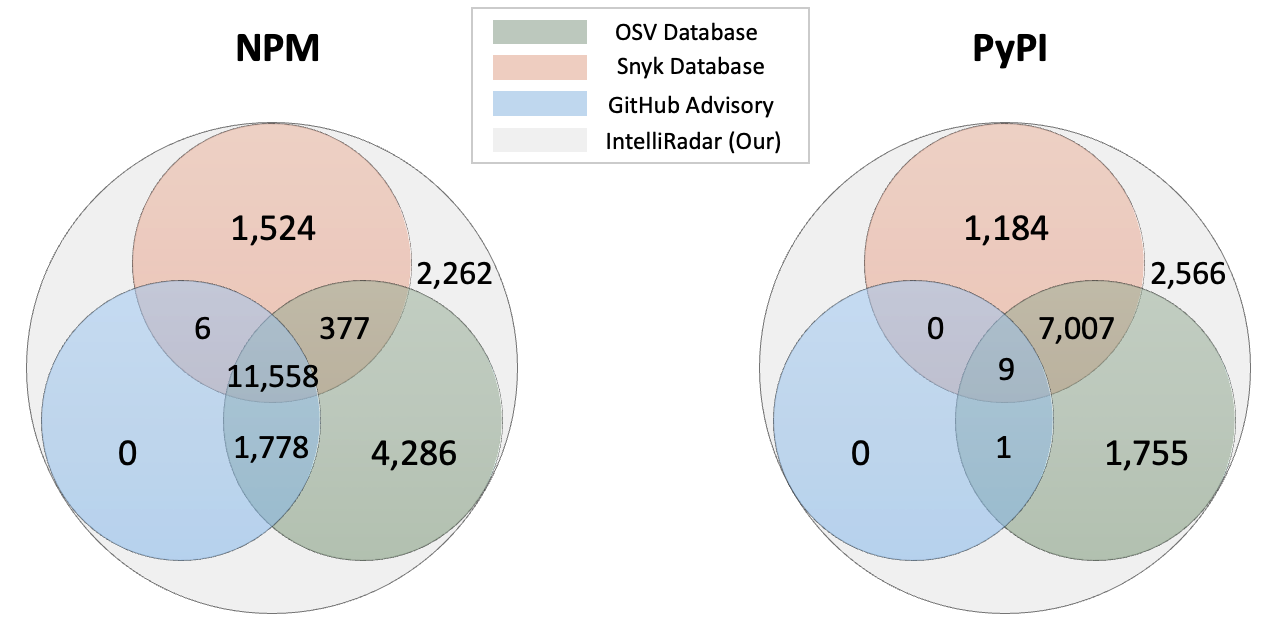}
  \caption{Comparison of Malicious Package Coverage in Different Databases}
  \label{fig:weien}
\end{figure}

In this section, we compare the data completeness between \tool and open-source databases including Snyk, OSV and GitHub Advisory. Figure~\ref{fig:weien} provides a detailed comparative analysis result. The results show that for PyPI malicious package intelligence, only 9 entries are common across all four databases, while OSV, Snyk, and \tool share 7,007 data points. For NPM malicious package data, 11,558 entries are common across all databases, primarily because NPM malicious intelligence is reported through GitHub Advisory, and other databases collect related intelligence based on this. 
%PyPI lacks such a unified mechanism, resulting in significant data discrepancies. 
%\wb{Notably, GitHub Advisory contains only 10 PyPI malicious packages, significantly fewer than other databases. This limitation stems from GitHub Advisory's focus on formal CVE assignments and its reliance on the Python Packaging Advisory Database, which maintains a conservative approach to advisory creation. Many PyPI malicious packages are detected and removed before formal advisories are issued, resulting in this substantial gap in coverage.}
GitHub Advisory contains only 10 PyPI malicious packages, significantly fewer than other databases. This disparity occurs because PyPI lacks a unified reporting mechanism, and GitHub Advisory primarily focuses on formal CVE assignments rather than malicious packages. Many PyPI malicious packages are detected and removed before formal advisories are issued, creating substantial coverage gaps in centralized databases.
%\tool encompasses all malicious package intelligence from OSV, Snyk, and GitHub Advisory for both PyPI and NPM ecosystems. Furthermore, through our comprehensive intelligence sources and LLM-based intelligence extraction framework, we uniquely identified an additional 2,262 NPM packages and 2,566 PyPI packages that were not present in any of the existing databases. This demonstrates the effectiveness of our approach in expanding the coverage of malicious package intelligence.

\tool includes data from OSV, Snyk, and GitHub Advisory to ensure comprehensive coverage, as these databases contain packages identified through internal detection tools and manual research not disclosed elsewhere. Beyond this data, our key contribution lies in extracting intelligence from other sources, identifying 17,759 NPM packages and 11,248 PyPI packages through LLM-based analysis. This includes 2,262 NPM packages and 2,566 PyPI packages not present in any existing structured databases, demonstrating our approach's effectiveness in discovering malicious packages from dispersed sources.
%Snyk database only allows access to only the first thirty pages of PyPI and NPM data, containing approximately 1,230 malicious packages. Some of its fields are incomplete. While the OSV database shows improvements in data completeness compared to Snyk, it still exhibits significant gaps in key features. Specifically, OSV provides accurate version information only for recently disclosed malicious packages, lacking version data for earlier ones. This absence of version information makes it difficult for downstream users to differentiate between malicious versions and regular releases.
%Table~\ref{tab:data-completeness-comparison} provides a detailed listing of malicious package information across different databases. Upon comparison, it shows \tool offers more comprehensive information on malicious packages. This includes specific version numbers, affected systems, attack vectors, and indicators of compromise (IOCs) found within the malicious code. 
% These detailed pieces of information greatly facilitate the analysis and utilization by downstream users and researchers.

% We extracted a total of 34,313 malicious package intelligence reports through collected intelligence sources. These included 12,522 PyPI packages and 21,791 NPM packages. Specifically, from 28,593 web pages across 24 intelligence sources, we extracted 11,069 PyPI and 17,691 NPM malicious packages intelligence from unstructured . 

The OSV database contains 17,999 NPM and 8,772 PyPI malicious packages. \tool covers all OSV database entries and identifies an additional 3,792 NPM and 3,750 PyPI malicious packages, representing 21.07\% and 42.74\% respectively.  Analysis of these unique data sources reveals that for NPM packages, Sonatype contributes 34.83\% of the intelligence, Phylum provides 19.29\%, with the remaining sources including QianXin, Medium, and Socket. For PyPI packages, Medium and Sonatype contribute 24.08\% and 16.26\% of the intelligence respectively, with additional sources including Phylum, Checkmarx, QianXin, Reddit, and TuxCare.

Comparison with the Snyk database shows that \tool shares 13,465 NPM and 8,200 PyPI malicious packages, while identifying 8,326 additional NPM packages (38.21\%) and 4,322 PyPI packages (34.52\%) through analyzing open-source web pages. Of these additional NPM packages, 25.51\% of the intelligence comes from GitHub and 16.90\% from Sonatype, with the remainder from Phylum and Socket. 
%For PyPI packages, the primary sources are Medium and Sonatype. These unique PyPI packages primarily employ Typosquatting attacks, with intelligence sources including Phylum, Checkmarx, and Twitter. 
For all PyPI packages collected by \tool from unstructured web content, the primary sources are Medium and Sonatype. Among the unique PyPI packages (those exclusively identified by \tool), the intelligence sources are primarily Phylum, Checkmarx, and Twitter, with these packages predominantly employing Typosquatting attacks.
Notably, \tool detected 749 packages on December 4, 2023, the same day they were released. The results demonstrate that LLM-based unstructured data analysis expands the coverage of existing databases. 
%This expansion is particularly significant as independent security practitioners often publish their findings through alternative channels not captured by traditional structured databases. By integrating these unstructured sources, \tool provides a broader view of malicious package information, enhancing the completeness of security intelligence in the open-source ecosystem.

To quantify the real-world impact of these uniquely identified malicious packages, we analyzed their downstream adoption through download statistics. We focused this analysis on PyPI packages, as PyPI officially provides comprehensive download records through "pypi.file\_downloads" in Google BigQuery, while no comparable data exists for NPM. Our analysis revealed that the 2,566 PyPI malicious packages uniquely identified by \tool were collectively downloaded 326,740 times before being detected. Among these, 1,427 packages were downloaded more than 100 times each.

\responsebox{
Response to RQ2: \tool is more comprehensive than well-known databases like OSV and Snyk. Additionally, it collected 4,828~(PyPI: 2,566 and NPM: 2,262) exclusive pieces of intelligence from all data sources.
}

\subsection{RQ3: Timeliness}
\label{sec:rq3}

\begin{figure}[]
  \centering
  \includegraphics[scale=0.20]{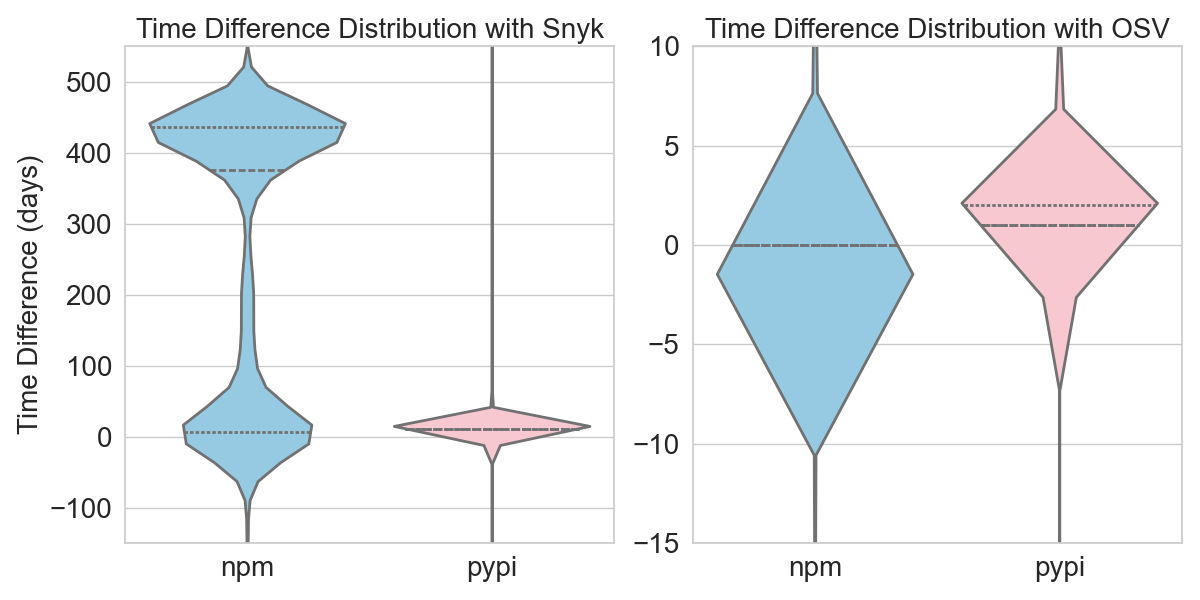}
  \caption{Time Intervals: IntelliRadar (Excluding OSV/Snyk) vs. OSV and Snyk}
  \label{fig:day_gap}
\end{figure}

%To validate how \tool outperforms existing malicious package databases in timeliness, we compared the dates when \tool and these databases recorded malicious packages.
To validate how \tool outperforms existing malicious package databases in timeliness, we compared the dates when \tool and these databases recorded malicious packages. To ensure a fair comparison, we specifically evaluated malicious packages collected by \tool from sources other than OSV and Snyk databases, comparing their detection timestamps against the same packages recorded in OSV and Snyk.

\textbf{Compared to the Snyk database.} For 10,314 NPM malicious packages, \tool recorded the intelligence earlier than Snyk, accounting for 76.6\%; among these, 6,598 packages were recorded 100.3 days earlier, with 49\% of packages recorded a year in advance. For PyPI malicious packages, 5,765 packages were recorded earlier than Snyk, accounting for 70.3\%, with 5,196 packages recorded more than a week earlier. Additionally, 566 malicious packages had consistent recording times, and only 189 packages (2.3\%) were recorded earlier by Snyk. The earlier detection capabilities of \tool stem from our comprehensive coverage of social media and unstructured websites from security companies, enabling us to discover the latest malicious packages immediately. For the packages that Snyk detected earlier, their internal proprietary detection tools identified these specific threats before public disclosure. These delays in Snyk are attributed to its dependency on established databases, GitHub security advisories, and manual research processes~\cite{snykvulndb}.

\begin{figure}[]
  \centering
  \includegraphics[scale=0.1]{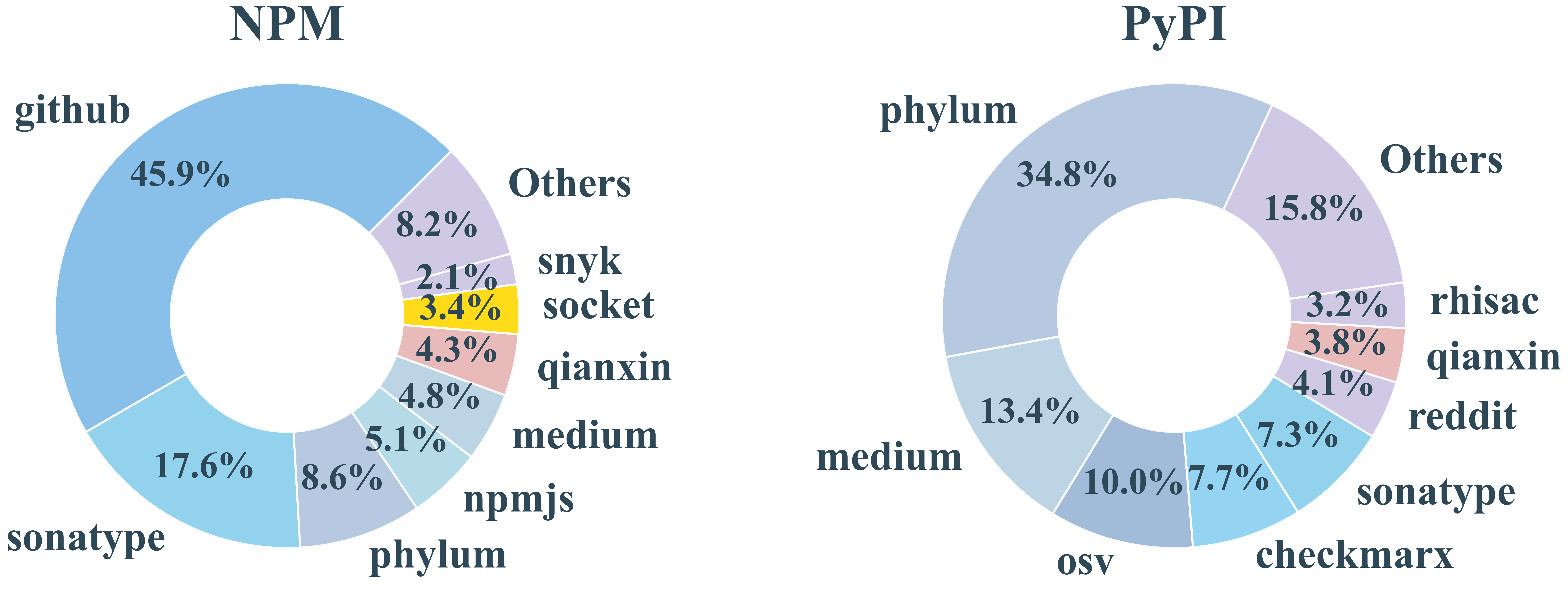}
  \caption{Distribution of Collected Malicious Intelligence Across Various Intelligence Sources}
  \label{fig:distribution_proportion}
\end{figure}

\textbf{Compared to the OSV database.} \tool and OSV have consistent recording times for 15,479 NPM packages, accounting for 86.0\% of the total. Additionally, \tool recorded 171 malicious packages earlier than OSV, primarily collected from Sonatype and Securityaffairs~\cite{securityaffairs_blog_ref}. For PyPI malicious packages, \tool recorded 4,711 earlier than OSV, with most differences within a week; 53.7\% were just one day earlier, mainly from Phylum. 511 packages were recorded on the same day by both \tool and OSV. Furthermore, we extracted PyPI package release times from Google's \textit{bigquery-public-data} and NPM package release times from the {NPM Registry}. Analysis shows that 4,533 NPM and 549 PyPI malicious package names were identified by \tool on the day of their release, indicating our method can swiftly collect relevant intelligence once malicious packages are publicly reported. OSV's delays are attributed to its aggregation model, which depends on multiple upstream databases~\cite{osvdata} and integrated detection tools~\cite{packageanalysis}.

To demonstrate the significance of timeliness of be cognitive to malicious package intelligence, we compared the time intervals of malicious package inclusion across different databases and the number of downloads during these intervals. Figure~\ref{fig:day_gap} shows the distribution of time differences between \textit{\tool-OSV} and \textit{\tool-Snyk} for malicious packages. Notably, 4,916 malicious packages were recorded in the Snyk database on 07/03/2023, while \tool identified these package names on 24/02/2023, through the Phylum intelligence source. Google Cloud data reveals that during these 11 days, these packages were downloaded 201,556 times, averaging 41 downloads per package. Further analysis indicates an uneven distribution of downloads, with the United States accounting for 49.2\% and China for 19.4\%. Among these, 107 malicious packages were downloaded 100 times each, with one package named \textit{studypong} downloaded 233 times. These data suggest that these packages were widely distributed and affected numerous users before being recorded by mainstream security databases. Our method enables early identification and warning before malicious packages become widespread, significantly reducing their lifecycle and minimizing their impact while protecting user safety.

\responsebox{
Response to RQ3: 73.9\% of \tool's intelligence is recorded earlier than the OSV database, and 57.9\% of PyPI intelligence is recorded earlier than the Snyk database, effectively shortening the lifecycle of malicious packages.
}

\subsection{RQ4: Source Distribution}
\label{sec:rq4}

We conducted a quantitative analysis of intelligence sources for malicious packages in the PyPI and NPM ecosystems. As shown in the figure, the PyPI ecosystem exhibits a diversified distribution pattern, with Phylum platform (34.8\%), Medium community (13.4\%), and OSV database (10.04\%) serving as the primary intelligence sources; in contrast, the NPM ecosystem demonstrates a highly centralized distribution, dominated by GitHub Advisory (45.85\%), Sonatype (17.56\%), and Phylum (8.55\%). Through in-depth analysis, we found that 32.4\% of malicious packages were documented by multiple data sources, while in the NPM ecosystem, 87.27\% of malicious packages were recorded by only a single source, primarily attributable to NPM's adoption of GitHub Advisory as the main channel for publishing malicious package intelligence. For packages uniquely identified by \tool, in the NPM ecosystem, these were primarily contributed by Phylum (23.61\%), GitHub (21.32\%), and Sonatype (21.28\%); PyPI malicious packages came from Sonatype (32.50\%), Medium (22.22\%), Qianxin (11.1\%), and Reddit, which provided 157 packages (8.77\%). Regarding intelligence timeliness, our data revealed significant differences among sources. In the PyPI ecosystem, Checkmarx and Phylum led in malicious package reporting times, accounting for 44.42\% and 23.25\% of reports, respectively. Specifically, Checkmarx issued alerts 3.02 days earlier on average, while Phylum reported 2.77 days earlier. Similarly, in the NPM ecosystem, for packages documented by multiple sources, GitHub and Phylum demonstrated superior timeliness, accounting for 40.35\% and 17.67\% of early reports. GitHub Advisory identified threats 15.29 days earlier on average compared to other intelligence sources, while Phylum reported 6.62 days earlier. These time differences reflect variations in threat identification capabilities and information dissemination mechanisms among intelligence sources, providing valuable reference for building efficient security response systems.

\responsebox{
Response to RQ4: Key intelligence sources differ between ecosystems - PyPI mainly relies on Phylum, Medium and OSV while NPM centralizes around GitHub Advisory. For timeliness, Checkmarx and Phylum lead in PyPI (reporting 3.02 and 2.77 days earlier), while GitHub dominates NPM Intelligence.
}

\subsection{RQ5: Usability}
\label{sec:rq5}

\begin{table*}[t]
\centering
\caption{Comparison of \tool Downstream Mirror Retention and Inclusion Times in Other Databases}
\footnotesize
\renewcommand{\arraystretch}{1.1}
\setlength{\tabcolsep}{4pt}
% 删除这行（未使用）
% \newcolumntype{C}[1]{>{\centering\arraybackslash}m{#1}}
% 用resizebox替代adjustbox（功能完全相同）
\resizebox{0.8\textwidth}{!}{%
\begin{tabular}{|c|c|c|c|c|c|c|c|c|c|c|c|}  % 修复列定义：c@{}| 改为 c|
\hline
\textbf{Package Name} & \textbf{Versions} & \multicolumn{5}{c|}{\textbf{PyPI Mirrors}} & \multicolumn{4}{c|}{\textbf{Database}} & \textbf{Downloads} \\
\cline{3-11}
 &  & \textbf{Tsinghua} & \textbf{Tencent} & \textbf{Aliyun} & \textbf{Douban} & \textbf{BFSU} & \textbf{OSV} & \textbf{Snyk} & \textbf{GitHub Advisory} & \textbf{\tool (Our)} &  \\
\hline
1inch & 8.6 - 8.9 & - & \checkmark & - & \checkmark & - & \ding{55} & 10/10/22 (+3) & \ding{55} & 07/10/22 & 110 \\
\hline
libcontroltoolver & 4.86 & - & \checkmark & - & \checkmark & - & 26/02/23 (+0) & 07/03/23 (+9) & \ding{55} & 26/02/23 & 62 \\
\hline
matplotlyib & 1.0.0 & \checkmark & \checkmark & - & - & \checkmark & \ding{55} & 29/03/23 (+31) & \ding{55} & 26/02/23 & 64 \\
\hline
pipcryptov4 & 1.0.0 & - & \checkmark & - & \checkmark & - & 25/06/24 \textcolor{red}{(+266)} & 05/10/23 (+2) & \ding{55} & 03/10/23 & 301 \\
\hline
libideeee & 1.0.0 & \checkmark & \checkmark & - & - & - & 25/06/24 \textcolor{red}{(+294)} & 07/09/23 (+2) & \ding{55} & 05/09/23 & 74 \\
\hline
ethereum2 & 2.8.4, 2.8.6, 2.8.9 & \checkmark & \checkmark & - & - & - & \ding{55} & 17/12/23 \textcolor{red}{(+436)} & \ding{55} & 07/10/22 & 331 \\
\hline
gkjzjh146 & 1.3 & \checkmark & \checkmark & - & - & - & 14/05/23 \textcolor{red}{(+226)} & \ding{55} & \ding{55} & 30/09/22 & 638 \\
\hline
httprequesthub & 2.31.0 - 2.31.4 & - & \checkmark & - & - & \checkmark & 25/06/24 \textcolor{red}{(+186)} & 25/12/23 (+3) & \ding{55} & 22/12/23 & 496 \\
\hline
logic2 & 0.1.4 & \checkmark & - & - & - & - & \ding{55} & 29/03/23 \textcolor{red}{(+104)} & \ding{55} & 15/12/22 & 520 \\
\hline
pipsqlite3liberyV2 & 1.1.0 & - & - & - & - & - & \ding{55} & 22/05/23 \textcolor{red}{(+104)} & \ding{55} & 15/05/23 & 145 \\
\hline
flak7 & 4.5.2 & \checkmark & - & - & - & - & \ding{55} & 07/09/22 (+5) & \ding{55} & 02/09/22 & 18 \\
\hline
simpeljson & 4.5.2 & \checkmark & - & - & - & - & 11/02/23 & 20/06/23 \textcolor{red}{(+129)} & \ding{55} & 11/02/23 & 67 \\
\hline
pyward & 3.0 & \checkmark & - & - & - & - & \ding{55} & 08/09/23 (+1) & \ding{55} & 07/09/23 & 559 \\
\hline
studypong & 5.66, 7.16, 8.22, 10.45 & \checkmark & - & - & - & - & 25/02/23 (+1) & 07/03/23 (+11) & \ding{55} & 24/02/23 & 264 \\
\hline
reqkests & 2.28.1 & \checkmark & - & - & - & - & \ding{55} & 11/12/22 (+2) & \ding{55} & 09/12/22 & 143 \\
\hline
beautiflulsoup & 1.0.0 & \checkmark & - & - & - & \checkmark & \ding{55} & 29/03/24 (+1) & \ding{55} & 28/03/24 & 76 \\
\hline
pycryptdome & 4.4.2 & - & - & - & - & - & \ding{55} & \ding{55} & \ding{55} & 25/08/22 & 63 \\
\hline
1337z & 4.4.7 & \checkmark & - & - & - & - & \ding{55} & \ding{55} & \ding{55} & 31/08/22 & 176 \\
\hline
urllib7 & 1.26.12 & \checkmark & - & - & - & - & \ding{55} & \ding{55} & \ding{55} & 15/12/22 & 69 \\
\hline
urllib12 & 1.26.12, 1.30.0 & \checkmark & - & - & - & - & \ding{55} & \ding{55} & \ding{55} & 15/12/22 & 296 \\
\hline
\end{tabular}%
}
\label{table:new_packages}
\begin{minipage}{\textwidth}
\footnotesize
\textit{The symbol \textbf{\checkmark} indicates the presence of the malicious package in the mirror, while '-' indicates its absence. The symbol \textbf{\ding{55}} indicates missing intelligence. The notation \textbf{(+n)} shows the number of days the package was included later than in our \tool database. \textbf{Downloads} indicate the number of times the malicious package has been downloaded.}
\end{minipage}
\end{table*}

% In this section, we demonstrate the usefulness of our collection malicious package intelligence by scanning and reporting the persisted malicious package intelligence in different mirrors of package manager registries.

In the context of package management ecosystems, the rapid propagation of malicious packages from primary repositories to downstream mirrors presents a significant security challenge. When a malicious contributor uploads a malicious package to \textit{pypi.org}, it swiftly synchronizes across various mirror sources. Due to the varied synchronization mechanisms (full or incremental) employed by different mirrors, downstream mirrors receive no formal notification when malicious packages are removed from pypi.org, potentially resulting in the prolonged persistence of these threats within mirror repositories.

\begin{figure}[]
  \centering
  \includegraphics[scale=0.19]{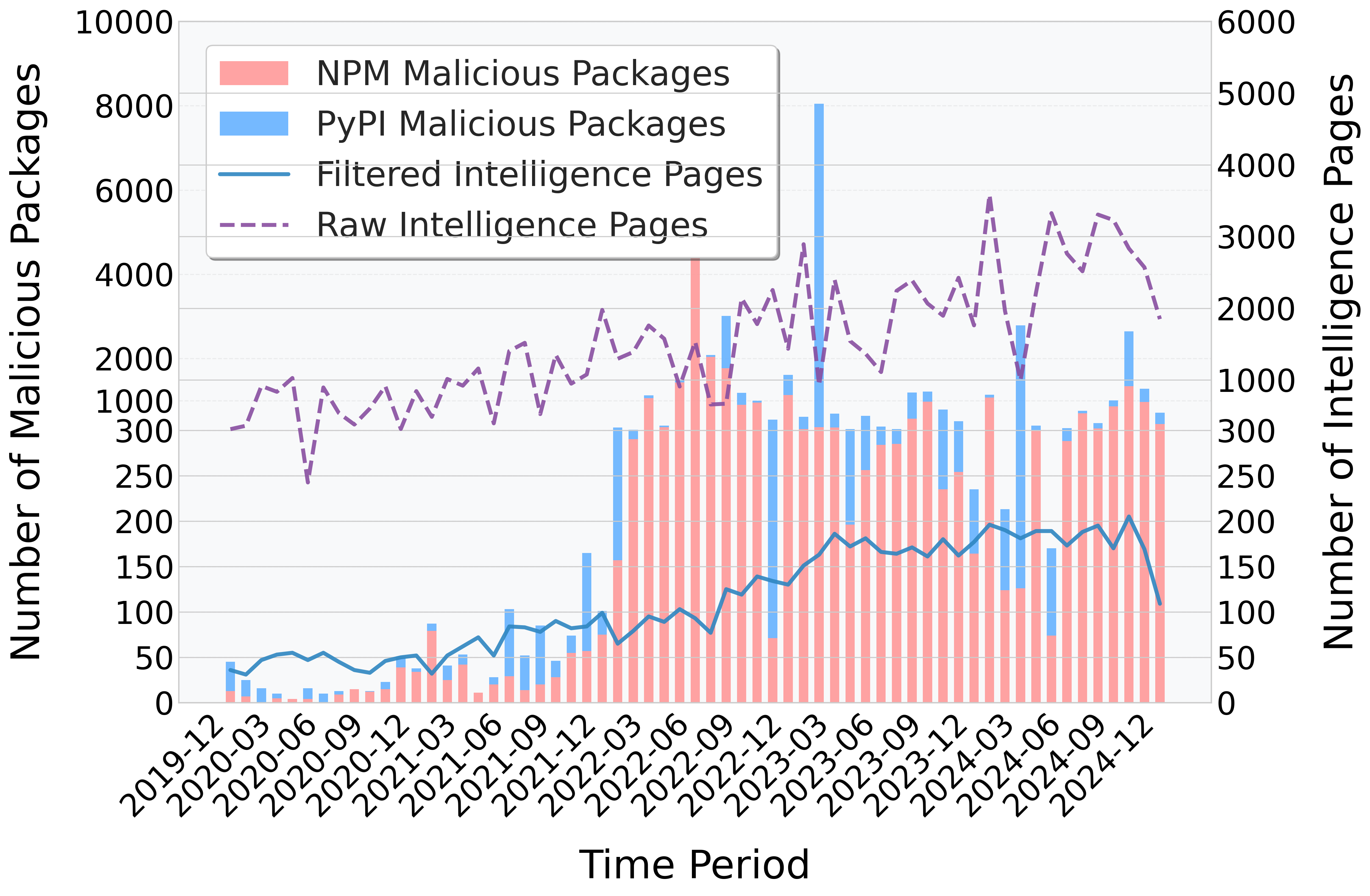}
  \caption{Monthly Analysis of Supply Chain Security Intelligence Sources and Malicious Packages}
  \label{fig:timelines_distribution}
\end{figure}

To this end, we scanned the main downstream PyPI mirrors~\cite{guo2023empirical} using the intelligence database. As of 01/05/2025, we discovered 1,981 malicious packages across various PyPI mirrors, specifically 781 in the Tsinghua mirror, 548 in the Tencent mirror, 395 in the Douban mirror, 21 in the BFSU mirror, 132 in the Huawei mirror, and 104 in the Aliyun mirror. Among these malicious packages, 54.7\% were uniquely identified in our database. These malicious packages had been downloaded over 86,240 times cumulatively, with the most severe case being the ethereum2 package, which had been downloaded 436 times before our identification and was identified 436 days earlier than other security databases such as Snyk. Table ~\ref{table:new_packages} presents the most impactful cases, highlighting the advantage of our tool in early identification.

In response to these findings, we sent batch emails to downstream mirror maintainers providing data on the detected malicious packages. We received confirmation replies from Tencent, Douban, and Tsinghua. By comparing the package status before and after our notification, we verified that the administrators had removed all the malicious packages. For real-time deployment data, we also issue monthly reports to downstream mirrors to ensure the security of the entire ecosystem.

Beyond these practical applications, we also evaluated the financial viability of \tool through a comprehensive cost analysis. We evaluated the costs of extracting malicious package information using LLMs. The initial webpage text from 24 data sources contained over 17 million tokens. After text filtering, this was reduced to 7.07 million tokens, a reduction of 58.4\%. The entire process cost \$93.8, We extracted a total of 34,313 pieces of intelligence. The average cost per piece of intelligence was \$0.003. This approach proved cost-effective, and we anticipate further cost reductions with future updates from OpenAI and open-source LLMs. 
%In our analysis of token consumption by intelligence sources, Phylum emerged as the primary contributor, accounting for 17.4\% of token usage and providing the highest number of malicious package entities. Other significant sources such as Sonatype and Tuxcare accounted for 10.5\% and 8.7\% of token consumption, respectively, indicating their roles as major providers of information. Particularly, platforms like Sonatype, Checkpoint, Checkmarx, and Medium are not only vital tools for open-source software monitoring but also primary sources of open-source intelligence, playing a crucial role in the identification and analysis of security threats.

%As shown in Figure \ref{fig:timelines_distribution}, we conducted a monthly distribution analysis of software supply chain security intelligence sources and malicious packages. The results demonstrate the significant intelligence extraction capability of \tool, averagely, it filters out 130 critical supply chain security intelligence pages from 2,371 original webpages monthly. Analysis of these high-quality intelligence pages identifies an average of 650 malicious packages per month from PyPI and NPM platforms, validating its effectiveness in threat \wb{identification}. With a monthly operational cost of only 7 USD dollars. The system exhibits exceptional cost-effectiveness in large-scale intelligence gathering. Notably, the data reveals a surge in malicious packages since 2022, indicating an escalating threat to software supply chain security.

As shown in Figure \ref{fig:timelines_distribution}, we conducted monthly distribution analysis of intelligence sources and malicious packages. \tool filters out 130 critical intelligence pages from 2,371 original webpages monthly, identifying an average of 650 malicious packages per month from PyPI and NPM platforms. With monthly operational costs of only 7 USD, the system demonstrates exceptional cost-effectiveness. The data reveals a surge in malicious packages since 2022, indicating escalating software supply chain threats.

\responsebox{
Response to RQ5: \tool identified and confirmed 1,981 malicious packages in downstream mirrors, with an intelligence cost of only \$0.003. Continuous monitoring for one month costs just \$7. As open-source LLMs continue to improve, these costs are expected to decrease further.
}

%% file: body/discussion.tex
\section{Discussion}

% \subsection{XXX title}

% \wenboreview{I don't know how to write this section}

% \subsection{Implication}

\noindent $\bullet$ \textbf{Intelligence Collection Beyond PyPI and NPM.}
In this study, we successfully collected a large amount of malicious package intelligence related to PyPI and NPM, as well as 928 other types of threat intelligence, covering component vulnerabilities, CVE vulnerabilities, and malicious package information from other package managers. It is worth noting that we discovered malicious components Prettier in the \textit{VSCode} Marketplace~\cite{prettier_vscode}, which uses typesquatting for attacks~\cite{prettier_vscode}. We also identified the \textit{XZ Utils} software supply chain attack case. Our approach can quickly adapt to the intelligence requirements of various package manager platforms and effectively expand to new security domains, providing the security community with a more timely automated intelligence discovery. 

\noindent $\bullet$ \textbf{Data Poisoning Attacks.} To combat potential data poisoning attacks, our framework implements multiple defensive mechanisms across the intelligence processing pipeline. During source identification, we meticulously screen and validate intelligence sources to ensure strong correlation with package management ecosystems, mitigating the risk of contaminated data. The intelligence aggregation mechanism employs a voting system to cross-validate entities from multiple independent sources, minimizing the impact of poisoned data. The intelligence extraction process strengthens defense through a three-step CoT approach (entity extraction, relationship alignment, and verification) that identifies and filters inconsistent or suspicious intelligence. The effectiveness is validated by experimental results: among 34,313 pieces of intelligence collected, 29,485 (85.9\%) were corroborated against authoritative security databases (OSV, Snyk, GitHub Advisory). This high correspondence rate demonstrates that our multilayered defense strategy successfully maintains the integrity and reliability of collected intelligence while effectively countering data poisoning attempts.

%% file: body/related.tex
\section{Related Works}

\subsection{Open Source Intelligence Analysis}

Open-source intelligence (OSINT) collects information through public channels, widely used in cybersecurity. Researchers have developed automated systems to gather threat intelligence from various sources, including the internet~\cite{mulwad2011extracting}, social media~\cite{sabottke2015vulnerability}, developer communities~\cite{purba2023extracting}, security forums, and code repositories~\cite{neil2018mining, bouwman2022helping, huang2021hackerrank}. 
Recent work has applied LLMs for security intelligence analysis~\cite{guo2025automated}, with CTIKG~\cite{huang2024ctikg} using LLMs for knowledge graph construction from cyber threat intelligence, and Høst et al.~\cite{hostconstructing} constructing knowledge graphs from vulnerability descriptions in the NVD. In software engineering, extensive research has focused on named entity recognition (NER) and relation extraction (RE) from unstructured text sources. Studies have extracted software entities from tweets~\cite{zhang2022benchmarking}, Stack Overflow posts~\cite{huang2023api, huo2022arclin}, and technical documents~\cite{nguyen2023software}, with recent work exploring self-evolving software intelligence management~\cite{liu2025evolaris}. These approaches have been applied to API mention resolution~\cite{huang2023api}, crash solutions~\cite{zhang2022benchmarking}, library recognition~\cite{zhang2022benchmarking}, and version incompatibility detection~\cite{zhao2023knowledge}. However, malicious package intelligence extraction presents unique challenges distinct from traditional software entity recognition. \tool requires: (1) distinguishing malicious package names from benign packages mentioned in the same text (where benign packages are often the legitimate targets that malicious packages attempt to mimic), (2) identifying the specific ecosystem (npm vs PyPI) for each package, and (3) performing semantic-level understanding to extract complex attack vectors and indicators of compromise.

\subsection{Software Supply Chain Security}

Software supply chain security faces severe challenges, with package management systems becoming hotspots for malicious activities. Attackers inject malicious code through methods like code tampering, dependency confusion, and typesquatting, leading to data leakage and system damage. DL Vu et al. reveal various attack methods and anti-detection techniques~\cite{vu2023bad, duan2020towards, gu2023investigating}. To address these threats, researchers have employed strategies ranging from static and dynamic analysis to machine learning. The MalOSS framework proposed by Duan et al.~\cite{duan2020towards} demonstrates similarities in malicious behaviors across different languages through metadata and API call sequence analysis. Huang et al. utilize graph-based behavior modeling to detect malicious packages~\cite{huang2024spiderscan}. Zhang et al.\cite{zhang2023malicious} utilize deep learning to understand malicious software's behavioral semantics, while Liang et al.\cite{liang2023needle} demonstrate the effectiveness of anomaly analysis algorithms. Additionally, GitHub, as the largest source code repository, may serve as a channel for distributing malicious code, with inconsistencies between distributed packages and source code potentially indicating malicious injection~\cite{vu2021lastpymile}. However, existing detection methods face a critical limitation: detection results are scattered across security blogs and unstructured webpages, failing to reach downstream users effectively. Even with numerous detection tools, malicious packages persist because intelligence remains fragmented. \tool addresses this through a fundamentally different approach—systematically analyzing malicious package intelligence from dispersed sources, covering both historical and newly reported packages.

%% file: body/conclusion.tex
\section{Conclusion}

% Package managers in the open-source ecosystem, such as PyPI and NPM, suffer from the absence of a comprehensive and current intelligence database, resulting in prolonged existence of malicious packages. To address this, we developed \tool, a new intelligence collecting system for package managers. It uses extensive data collection from multiple sources and large language models for enhanced intelligence extraction. Our approach gathered 34,313 pieces of intelligence about NPM and PyPI from 24 sources, providing more unique and detailed information than existing databases. It detects malicious packages earlier, reducing their lifecycle and download frequency. The method is cost-effective at \$0.003 per piece of intelligence, improving malware detection efficiency and strengthening open-source software supply chain security. Our open source dataset and supplementary datasets are available online~\cite{website}.

% In the open-source ecosystem, package managers like PyPI and NPM lack up-to-date intelligence databases, allowing malicious packages to persist. To address this issue, we developed \tool, a system that leverages multi-source data collection and LLMs to gather 34,313 pieces of intelligence about NPM and PyPI from 24 sources, offering unique and detailed insights. The system enables earlier detection of malicious packages, reducing their impact, with a cost-effective rate of \$0.003 per intelligence item. This work presents an efficient and economical solution to enhance open-source software supply chain security. 

In the open-source ecosystem, package managers like PyPI and NPM lack up-to-date intelligence databases, allowing malicious packages to persist. To address this, we developed \tool, a system leveraging multi-source data collection and LLMs to gather 34,313 pieces of intelligence from 24 sources. The system enables earlier detection of malicious packages, reducing their impact, at \$0.003 per intelligence item. This work presents an efficient solution to enhance open-source software supply chain security.

% \noindent $\bullet$ \textbf{Data Availability.} Our open source dataset and supplementary datasets are available online~\cite{website}.

\section*{Acknowledgment}

This work was supported by the National Key Research and Development Program of China (Grant No. 2024YFF0908000). This research is supported by the National Research Foundation, Singapore, and DSO National Laboratories under the AI Singapore Programme (AISG Award No: AISG4-GC-2023-008-1B); by the National Research Foundation Singapore and the Cyber Security Agency under the National Cybersecurity R\&D Programme (NCRP25-P04-TAICeN); and by the Prime Minister’s Office, Singapore under the Campus for Research Excellence and Technological Enterprise (CREATE) Programme.
Any opinions, findings and conclusions, or recommendations expressed in these materials are those of the author(s) and do not reflect the views of the National Research Foundation, Singapore, Cyber Security Agency of Singapore, Singapore.

%% file: ref.bib
@Misc{npm_stole_pwd,
note = {\url{https://www.bleepingcomputer.com/news/security/npm-pulls-malicious-package-that-stole-login-passwords/} [Accessed Apr 21, 2024]},
title = {NPM Pulls Malicious Package that Stole Login Passwords},
year = {Aug 21, 2019}
}

@Misc{npm_remote_trojans,
note = {\url{https://www.zdnet.com/article/malicious-npm-packages-caught-installing-remote-access-trojans/} [Accessed Apr 21, 2024]},
title = {Malicious npm packages caught installing remote access trojans},
year = {Dec. 1, 2020}
}

@Misc{pypi_malicious_trojans,
note = {\url{https://thehackernews.com/2023/12/116-malware-packages-found-on-pypi.html} [Accessed Apr 21, 2024]},
title = {116 Malware Packages Found on PyPI Repository Infecting Windows and Linux Systems},
year = {Dec 14, 2023}
}

@Misc{snyk_colorwed,
note = {\url{https://security.snyk.io/vuln?search=colorwed} [Accessed Apr 21, 2024]},
title = {Colorwed package information in Snyk Vulnerability Database},
year = {Dec 21,  2022}
}

@misc{huang2023surveyhallucinationlargelanguage,
      title={A Survey on Hallucination in Large Language Models: Principles, Taxonomy, Challenges, and Open Questions}, 
      author={Lei Huang and Weijiang Yu and Weitao Ma and Weihong Zhong and Zhangyin Feng and Haotian Wang and Qianglong Chen and Weihua Peng and Xiaocheng Feng and Bing Qin and Ting Liu},
      year={2023},
      eprint={2311.05232},
      archivePrefix={arXiv},
      primaryClass={cs.CL},
      url={https://arxiv.org/abs/2311.05232}, 
}

@inproceedings{guo2023empirical,
  title={An Empirical Study of Malicious Code In PyPI Ecosystem},
  author={Guo, Wenbo and Xu, Zhengzi and Liu, Chengwei and Huang, Cheng and Fang, Yong and Liu, Yang},
  booktitle={2023 38th IEEE/ACM International Conference on Automated Software Engineering (ASE)},
  pages={166--177},
  year={2023},
  organization={IEEE}
}

@Misc{colorwed_twitter,
note = {\url{https://twitter.com/LouiswLang/status/1595270411201441793/} [Accessed Apr 21, 2024]},
title = {More w4sp malware packages published to pypi},
year = {Nov 23, 2022}
}

@Misc{colorwed_jfrog,
note = {\url{https://twitter.com/JFrogSecurity/status/1595755792577200128/} [Accessed Apr 21, 2024]},
title = {Catching more npm malicious packages related to the ongoing W4SP infector campaign. },
year = {Nov 24, 2022}
}

@Misc{colorwed_phylum,
note = {\url{https://blog.phylum.io/w4sp-stealer-update-theyre-still-at-it/} [Accessed Apr 21, 2024]},
title = {W4SP Stealer Update—They’re Still At It},
year = {Dec 15, 2022}
}

@Misc{colorwed_sonatype,
note = {\url{https://blog.sonatype.com/malware-monthly-november-2022/} [Accessed Apr 21, 2024]},
title = {Malware Monthly - November 2022},
year = {Dec 15, 2022}
}

@Misc{snyk_database_ref,
note = {\url{https://security.snyk.io/vuln/} [Accessed Apr 21, 2024]},
title = {Snyk Vulnerability Database},
year = {Jan 25, 2024}
}

@Misc{osv_database_ref,
note = {\url{https://osv.dev/} [Accessed Apr 21, 2024]},
title = {A distributed vulnerability database for Open Source},
year = {Apr 01, 2022}
}

@Misc{advisory_database_ref,
note = {\url{https://github.com/advisories?query=type\&malware/} [Accessed Apr 21, 2024]},
title = {GitHub Advisory Database},
year = {Jun 8, 2022}
}

@Misc{securityaffairs_blog_ref,
note = {\url{https://securityaffairs.com/} [Accessed Apr 21, 2024]},
title = {Securityaffairs News},
year = {Jul 15, 2019}
}

@Misc{prettier_vscode,
note = {\url{https://www.bleepingcomputer.com/news/microsoft/vscode-marketplace-can-be-abused-to-host-malicious-extensions/} [Accessed Apr 21, 2024]},
title = {VSCode Marketplace can be abused to host malicious extensions},
year = {Jan 6, 2023}
}

@inproceedings{sabottke2015vulnerability,
  title={Vulnerability disclosure in the age of social media: Exploiting twitter for predicting $\{$Real-World$\}$ exploits},
  author={Sabottke, Carl and Suciu, Octavian and Dumitraș, Tudor},
  booktitle={24th USENIX Security Symposium (USENIX Security 15)},
  pages={1041--1056},
  year={2015}
}

@inproceedings{mulwad2011extracting,
  title={Extracting information about security vulnerabilities from web text},
  author={Mulwad, Varish and Li, Wenjia and Joshi, Anupam and Finin, Tim and Viswanathan, Krishnamurthy},
  booktitle={2011 IEEE/WIC/ACM International Conferences on Web Intelligence and Intelligent Agent Technology},
  volume={3},
  pages={257--260},
  year={2011},
  organization={IEEE}
}

@inproceedings{tong2016text,
  title={A text mining research based on LDA topic modelling},
  author={Tong, Zhou and Zhang, Haiyi},
  booktitle={International conference on computer science, engineering and information technology},
  pages={201--210},
  year={2016}
}

@inproceedings{tahir2018s,
  title={It's all in the name: Why some URLs are more vulnerable to typosquatting},
  author={Tahir, Rashid and Raza, Ali and Ahmad, Faizan and Kazi, Jehangir and Zaffar, Fareed and Kanich, Chris and Caesar, Matthew},
  booktitle={IEEE INFOCOM 2018-IEEE Conference on Computer Communications},
  pages={2618--2626},
  year={2018},
  organization={IEEE}
}

@inproceedings{vu2020typosquatting,
  title={Typosquatting and combosquatting attacks on the python ecosystem},
  author={Vu, Duc-Ly and Pashchenko, Ivan and Massacci, Fabio and Plate, Henrik and Sabetta, Antonino},
  booktitle={2020 ieee european symposium on security and privacy workshops (euros\&pw)},
  pages={509--514},
  year={2020},
  organization={IEEE}
}

@inproceedings{martino2023knowledge,
  title={Knowledge injection to counter large language model (LLM) hallucination},
  author={Martino, Ariana and Iannelli, Michael and Truong, Coleen},
  booktitle={European Semantic Web Conference},
  pages={182--185},
  year={2023},
  organization={Springer}
}

@article{qaiser2018text,
  title={Text mining: use of TF-IDF to examine the relevance of words to documents},
  author={Qaiser, Shahzad and Ali, Ramsha},
  journal={International Journal of Computer Applications},
  volume={181},
  number={1},
  pages={25--29},
  year={2018},
  publisher={Foundation of Computer Science}
}

@article{min2023recent,
  title={Recent advances in natural language processing via large pre-trained language models: A survey},
  author={Min, Bonan and Ross, Hayley and Sulem, Elior and Veyseh, Amir Pouran Ben and Nguyen, Thien Huu and Sainz, Oscar and Agirre, Eneko and Heintz, Ilana and Roth, Dan},
  journal={ACM Computing Surveys},
  volume={56},
  number={2},
  pages={1--40},
  year={2023},
  publisher={ACM New York, NY}
}

@article{wang2023gpt,
  title={Gpt-ner: Named entity recognition via large language models},
  author={Wang, Shuhe and Sun, Xiaofei and Li, Xiaoya and Ouyang, Rongbin and Wu, Fei and Zhang, Tianwei and Li, Jiwei and Wang, Guoyin},
  journal={arXiv preprint arXiv:2304.10428},
  year={2023}
}

@inproceedings{purba2023extracting,
  title={Extracting Actionable Cyber Threat Intelligence from Twitter Stream},
  author={Purba, Moumita Das and Chu, Bill},
  booktitle={2023 IEEE International Conference on Intelligence and Security Informatics (ISI)},
  pages={1--6},
  year={2023},
  organization={IEEE}
}

@article{huang2021hackerrank,
  title={HackerRank: Identifying key hackers in underground forums},
  author={Huang, Cheng and Guo, Yongyan and Guo, Wenbo and Li, Ying},
  journal={International Journal of Distributed Sensor Networks},
  volume={17},
  number={5},
  pages={15501477211015145},
  year={2021},
  publisher={SAGE Publications Sage UK: London, England}
}

@inproceedings{bouwman2022helping,
  title={Helping hands: Measuring the impact of a large threat intelligence sharing community},
  author={Bouwman, Xander and Le Pochat, Victor and Foremski, Pawel and Van Goethem, Tom and Ga{\~n}{\'a}n, Carlos H and Moura, Giovane CM and Tajalizadehkhoob, Samaneh and Joosen, Wouter and Van Eeten, Michel},
  booktitle={31st USENIX Security Symposium (USENIX Security 22)},
  pages={1149--1165},
  year={2022}
}

@inproceedings{neil2018mining,
  title={Mining threat intelligence about open-source projects and libraries from code repository issues and bug reports},
  author={Neil, Lorenzo and Mittal, Sudip and Joshi, Anupam},
  booktitle={2018 IEEE International Conference on Intelligence and Security Informatics (ISI)},
  pages={7--12},
  year={2018},
  organization={IEEE}
}

@inproceedings{gu2023investigating,
  title={Investigating package related security threats in software registries},
  author={Gu, Yacong and Ying, Lingyun and Pu, Yingyuan and Hu, Xiao and Chai, Huajun and Wang, Ruimin and Gao, Xing and Duan, Haixin},
  booktitle={2023 IEEE Symposium on Security and Privacy (SP)},
  pages={1578--1595},
  year={2023},
  organization={IEEE}
}

@inproceedings{liang2023needle,
  title={A Needle is an Outlier in a Haystack: Hunting Malicious PyPI Packages with Code Clustering},
  author={Liang, Wentao and Ling, Xiang and Wu, Jingzheng and Luo, Tianyue and Wu, Yanjun},
  booktitle={2023 38th IEEE/ACM International Conference on Automated Software Engineering (ASE)},
  pages={307--318},
  year={2023},
  organization={IEEE}
}

@article{zhang2023malicious,
  title={Malicious Package Detection in NPM and PyPI using a Single Model of Malicious Behavior Sequence},
  author={Zhang, Junan and Huang, Kaifeng and Chen, Bihuan and Wang, Chong and Tian, Zhenhao and Peng, Xin},
  journal={arXiv preprint arXiv:2309.02637},
  year={2023}
}

@article{duan2020towards,
  title={Towards measuring supply chain attacks on package managers for interpreted languages},
  author={Duan, Ruian and Alrawi, Omar and Kasturi, Ranjita Pai and Elder, Ryan and Saltaformaggio, Brendan and Lee, Wenke},
  journal={arXiv preprint arXiv:2002.01139},
  year={2020}
}

@inproceedings{vu2021lastpymile,
  title={Lastpymile: identifying the discrepancy between sources and packages},
  author={Vu, Duc-Ly and Massacci, Fabio and Pashchenko, Ivan and Plate, Henrik and Sabetta, Antonino},
  booktitle={Proceedings of the 29th ACM Joint Meeting on European Software Engineering Conference and Symposium on the Foundations of Software Engineering},
  pages={780--792},
  year={2021}
}

@inproceedings{vu2023bad,
  title={Bad Snakes: Understanding and Improving Python Package Index Malware Scanning},
  author={Vu, Duc-Ly and Newman, Zachary and Meyers, John Speed},
  booktitle={2023 IEEE/ACM 45th International Conference on Software Engineering (ICSE)},
  pages={499--511},
  year={2023},
  organization={IEEE}
}

@article{10.1093/cybsec/tyaa015,
    author = {Jacobs, Jay and Romanosky, Sasha and Adjerid, Idris and Baker, Wade},
    title = "{Improving vulnerability remediation through better exploit prediction}",
    journal = {Journal of Cybersecurity},
    volume = {6},
    number = {1},
    pages = {tyaa015},
    year = {2020},
    month = {09},
    issn = {2057-2085},
    doi = {10.1093/cybsec/tyaa015},
    url = {https://doi.org/10.1093/cybsec/tyaa015},
    eprint = {https://academic.oup.com/cybersecurity/article-pdf/6/1/tyaa015/33746021/tyaa015.pdf},
}

@article{ling2023improving,
  title={Improving open information extraction with large language models: A study on demonstration uncertainty},
  author={Ling, Chen and Zhao, Xujiang and Zhang, Xuchao and Liu, Yanchi and Cheng, Wei and Wang, Haoyu and Chen, Zhengzhang and Osaki, Takao and Matsuda, Katsushi and Chen, Haifeng and others},
  journal={arXiv preprint arXiv:2309.03433},
  year={2023}
}

@inproceedings{ladisa2023sok,
  title={Sok: Taxonomy of attacks on open-source software supply chains},
  author={Ladisa, Piergiorgio and Plate, Henrik and Martinez, Matias and Barais, Olivier},
  booktitle={2023 IEEE Symposium on Security and Privacy (SP)},
  pages={1509--1526},
  year={2023},
  organization={IEEE}
}

@inproceedings{okafor2022sok,
  title={Sok: Analysis of software supply chain security by establishing secure design properties},
  author={Okafor, Chinenye and Schorlemmer, Taylor R and Torres-Arias, Santiago and Davis, James C},
  booktitle={Proceedings of the 2022 ACM Workshop on Software Supply Chain Offensive Research and Ecosystem Defenses},
  pages={15--24},
  year={2022}
}

@article{ladisa2023journey,
  title={Journey to the Center of Software Supply Chain Attacks},
  author={Ladisa, Piergiorgio and Ponta, Serena Elisa and Sabetta, Antonino and Martinez, Matias and Barais, Olivier},
  journal={arXiv preprint arXiv:2304.05200},
  year={2023}
}

@article{zhou2024large,
  title={A Large-scale Fine-grained Analysis of Packages in Open-Source Software Ecosystems},
  author={Zhou, Xiaoyan and Liang, Feiran and Xie, Zhaojie and Lan, Yang and Niu, Wenjia and Liu, Jiqiang and Wang, Haining and Li, Qiang},
  journal={arXiv preprint arXiv:2404.11467},
  year={2024}
}

@article{zhou2024oss,
  title={OSS Malicious Package Analysis in the Wild},
  author={Zhou, Xiaoyan and Zhang, Ying and Niu, Wenjia and Liu, Jiqiang and Wang, Haining and Li, Qiang},
  journal={arXiv preprint arXiv:2404.04991},
  year={2024}
}

@article{huang2024donapi,
  title={DONAPI: Malicious NPM Packages Detector using Behavior Sequence Knowledge Mapping},
  author={Huang, Cheng and Wang, Nannan and Wang, Ziyan and Sun, Siqi and Li, Lingzi and Chen, Junren and Zhao, Qianchong and Han, Jiaxuan and Yang, Zhen and Shi, Lei},
  journal={arXiv preprint arXiv:2403.08334},
  year={2024}
}

@inproceedings{li2023malwukong,
  title={MalWuKong: Towards Fast, Accurate, and Multilingual Detection of Malicious Code Poisoning in OSS Supply Chains},
  author={Li, Ningke and Wang, Shenao and Feng, Mingxi and Wang, Kailong and Wang, Meizhen and Wang, Haoyu},
  booktitle={2023 38th IEEE/ACM International Conference on Automated Software Engineering (ASE)},
  pages={1993--2005},
  year={2023},
  organization={IEEE}
}

@article{zahan2024shifting,
  title={Shifting the Lens: Detecting Malware in npm Ecosystem with Large Language Models},
  author={Zahan, Nusrat and Burckhardt, Philipp and Lysenko, Mikola and Aboukhadijeh, Feross and Williams, Laurie},
  journal={arXiv preprint arXiv:2403.12196},
  year={2024}
}

@Misc{snykweb,
note = {\url{https://snyk.io/product/open-source-security-management/} [Accessed Apr 30, 2024]},
title = {Open source risk management made for developers.},
year = {2024}
}

@Misc{BlackDuckweb,
note = {\url{https://www.synopsys.com/software-integrity/software-composition-analysis-tools/binary-analysis.html} [Accessed Apr 30, 2024]},
title = {Black Duck Binary Analysis: Identify open source supply chain risks even when you don't have access to the code.},
year = {2024}
}

@Misc{OWASPweb,
note = {\url{https://owasp.org/www-project-dependency-check/} [Accessed Apr 30, 2024]},
title = {OWASP Dependency-Check.},
year = {2024}
}

@Misc{dependabotweb,
note = {\url{https://github.com/dependabot} [Accessed Apr 30, 2024]},
title = {Dependabot: Automated dependency updates built into GitHub.},
year = {2024}
}

@Misc{nvdweb,
note = {\url{https://nvd.nist.gov/} [Accessed Apr 30, 2024]},
title = {NATIONAL VULNERABILITY DATABASE.},
year = {2024}
}

@misc{website,
author = {},
title = {IntelliRadar},
howpublished = {\url{https://sites.google.com/view/intelliradar/home}},
month = {},
year = {2024},
note = {(Accessed on 04/30/2024)}
}

@inproceedings{ohm2020backstabber,
  title={Backstabber’s knife collection: A review of open source software supply chain attacks},
  author={Ohm, Marc and Plate, Henrik and Sykosch, Arnold and Meier, Michael},
  booktitle={Detection of Intrusions and Malware, and Vulnerability Assessment: 17th International Conference, DIMVA 2020, Lisbon, Portugal, June 24--26, 2020, Proceedings 17},
  pages={23--43},
  year={2020},
  organization={Springer}
}

@Misc{snykvulndb,
note = {\url{https://docs.snyk.io/scan-with-snyk/snyk-open-source/manage-vulnerabilities/snyk-vulnerability-database} [Accessed Jan 13, 2025]},
title = {Snyk Vulnerability Database},
year = {2024}
}

@Misc{osvdata,
note = {\url{https://google.github.io/osv.dev/data/} [Accessed Jan 13, 2025]},
title = {OSV Data Sources},
year = {2024}
}

@Misc{packageanalysis,
note = {\url{https://github.com/ossf/package-analysis} [Accessed Jan 13, 2025]},
title = {Package Analysis: Open source package analysis},
year = {2024}
}

@Misc{npmregistry,
note = {\url{https://registry.npmjs.org/angilarjs/8.7.6} [Accessed Jul 05, 2025]},
title = {npm Registry API},
year = {2024}
}

@Misc{pypistats,
note = {\url{https://pypistats.org/api/} [Accessed Jul 05, 2025]},
title = {PyPI Stats API},
year = {2024}
}

@Misc{pypibigquery,
note = {\url{https://bigquery.cloud.google.com/table/bigquery-public-data:pypi.downloads} [Accessed Jul 05, 2025]},
title = {PyPI Downloads Public Dataset on Google BigQuery},
year = {2024}
}

@inproceedings{huang2024ctikg,
  title={Ctikg: Llm-powered knowledge graph construction from cyber threat intelligence},
  author={Huang, Liangyi and Xiao, Xusheng},
  booktitle={First Conference on Language Modeling},
  year={2024}
}

@inproceedings{hostconstructing,
  title={Constructing a Knowledge Graph from Textual Descriptions of Software Vulnerabilities in the National Vulnerability Database},
  author={H{\o}st, Anders M{\o}lmen and Lison, Pierre and Moonen, Leon},
  booktitle={The 24rd Nordic Conference on Computational Linguistics}
}

@article{huang2023api,
  title={Api entity and relation joint extraction from text via dynamic prompt-tuned language model},
  author={Huang, Qing and Sun, Yanbang and Xing, Zhenchang and Yu, Min and Xu, Xiwei and Lu, Qinghua},
  journal={ACM transactions on software engineering and methodology},
  volume={33},
  number={1},
  pages={1--25},
  year={2023},
  publisher={ACM New York, NY}
}

@inproceedings{huo2022arclin,
  title={ARCLIN: automated API mention resolution for unformatted texts},
  author={Huo, Yintong and Su, Yuxin and Zhang, Hongming and Lyu, Michael R},
  booktitle={Proceedings of the 44th International Conference on Software Engineering},
  pages={138--149},
  year={2022}
}

@inproceedings{nguyen2023software,
  title={Software entity recognition with noise-robust learning},
  author={Nguyen, Tai and Di, Yifeng and Lee, Joohan and Chen, Muhao and Zhang, Tianyi},
  booktitle={2023 38th IEEE/ACM International Conference on Automated Software Engineering (ASE)},
  pages={484--496},
  year={2023},
  organization={IEEE}
}

@inproceedings{zhang2022benchmarking,
  title={Benchmarking library recognition in tweets},
  author={Zhang, Ting and Chandrasekaran, Divya Prabha and Thung, Ferdian and Lo, David},
  booktitle={Proceedings of the 30th IEEE/ACM international conference on program comprehension},
  pages={343--353},
  year={2022}
}

@inproceedings{zhao2023knowledge,
  title={Knowledge-based version incompatibility detection for deep learning},
  author={Zhao, Zhongkai and Kou, Bonan and Ibrahim, Mohamed Yilmaz and Chen, Muhao and Zhang, Tianyi},
  booktitle={Proceedings of the 31st ACM Joint European Software Engineering Conference and Symposium on the Foundations of Software Engineering},
  pages={708--719},
  year={2023}
}

@inproceedings{huang2024spiderscan,
  title={SpiderScan: Practical detection of malicious NPM packages based on graph-based behavior modeling and matching},
  author={Huang, Yiheng and Wang, Ruisi and Zheng, Wen and Zhou, Zhuotong and Wu, Susheng and Ke, Shulin and Chen, Bihuan and Gao, Shan and Peng, Xin},
  booktitle={Proceedings of the 39th IEEE/ACM International Conference on Automated Software Engineering},
  pages={1146--1158},
  year={2024}
}

@inproceedings{host2023constructing,
  title={Constructing a Knowledge Graph from Textual Descriptions of Software Vulnerabilities in the National Vulnerability Database},
  author={H{\o}st, Anders and Lison, Pierre and Moonen, Leon},
  booktitle={Proceedings of the 24th Nordic Conference on Computational Linguistics (NoDaLiDa)},
  pages={386--391},
  year={2023}
}

@inproceedings{liu2025evolaris,
  title={Evolaris: A Roadmap to Self-evolving Software Intelligence Management},
  author={Liu, Chengwei and Guo, Wenbo and Zhang, Yuxin and Wang, Limin and Chen, Sen and Bu, Lei and Liu, Yang},
  booktitle={International Conference on Engineering of Complex Computer Systems},
  pages={515--520},
  year={2025},
  organization={Springer}
}

@inproceedings{guo2025automated,
  title={Automated Environment Extraction for Malicious Package Validation: Leveraging Threat Intelligence},
  author={Guo, Wenbo and Wang, Limin and Zhang, Yiran and Xu, Zhengzi and Wu, Jiahui},
  booktitle={Proceedings of the 33rd ACM International Conference on the Foundations of Software Engineering},
  pages={1756--1759},
  year={2025}
}
